\documentclass[epj]{svjour}
\usepackage{graphicx}

\usepackage{color}
\usepackage{amssymb}
\usepackage{bbm}
\usepackage[intlimits]{amsmath}
\usepackage{widetext}
\newcommand{\OperatorOne}{\mathrm{1\hspace{-0.13cm}1}} 
\newcommand{\sh}[1]{#1\hskip-7pt \diagup}
\newcommand{\Sh}[1]{#1\hskip-10pt \diagup}
\newcommand{\tr}{\textrm{tr}}

\begin{document}
\title{Beyond Rainbow-Ladder in bound state equations}
\author{Walter Heupel\inst{1} \and Tobias Goecke\inst{1}
\and Christian S. Fischer \inst{1}
}                     
\institute{Institute for Theoretical Physics, Justus-Liebig-University Gie\ss en} 
\date{Received: date / Revised version: date}
\abstract{
In this work we devise a new method to study quark anti-quark interactions
beyond simple ladder-exchange that yield massless pions in the chiral limit. 
The method is based on the requirement to have a representation of the quark-gluon 
vertex that is explicitly given in terms of quark dressings functions. We outline 
a general procedure to generate the Bethe-Salpeter kernel for a given vertex 
representation. Our method allows not only the identification of the mesons'
masses but also the extraction of their Bethe-Salpeter wave functions exposing
their internal structure. We exemplify our method with vertex models that are of 
phenomenological interest.
} %end of abstract
\maketitle
\section{Introduction}
\label{intro}
Understanding the spectrum of light and heavy hadrons is an important task 
on our way towards a full understanding of QCD. In order to identify states 
that can be accounted for as quark-antiquark bound systems and separate them 
from more complex ones such as tetraquarks, meson molecules or glueballs one 
needs to develop a framework that makes contact to the details of the underlying 
quark-gluon interaction. Lattice QCD is one such approach, the functional method
using Dyson-Schwinger \linebreak equations (DSEs) and Bethe-Salpeter equations (BSEs) is another. 

In the latter approach, the construction of an approximation scheme that yields 
an interaction consistent with chiral symmetry and its breaking patterns is a 
necessary requirement for the description of light mesons. Only then, the Goldstone
boson nature of the pseudoscalar bound states are preserved resulting in a massless 
pion in the chiral limit \cite{McKay:1989rk,Munczek:1994zz,Maris:1997hd}.
This requirement can most easily be met with the rainbow-ladder
truncation which has been widely applied for QCD phenomenology
\cite{Maris:2003vk,Maris:2005tt,Goecke:2011pe,Eichmann:2009qa}. 
This truncation has, however, limitations. These become visible for 
exited states \cite{Eichmann:2008ae,Qin:2011xq}, states with finite 
width, or mesons with axial-vector or scalar quantum numbers,
where the rainbow-ladder approach does not provide results in agreement
with experiment. On a fundamental level, going beyond simple models for 
the quark-gluon interaction requires a dynamical treatment of the Yang-Mills 
sector of QCD as well as a treatment of the quark-gluon vertex that
includes beyond rainbow-ladder structures \cite{Fischer:2003rp}.

There have been many efforts to go beyond rainbow-ladder. One promising 
route is to use explicit diagrammatic approximations to the DSE of the 
quark-gluon vertex \cite{Bender:1996bb,Watson:2004kd,Bhagwat:2004hn,%
Matevosyan:2006bk,Alkofer:2008tt,Fischer:2007ze,Fischer:2009jm,Fischer:2008wy}.
This allows the explicit study of the effects of the gluon self-interaction
\cite{Fischer:2009jm} as well as pion cloud effects \cite{Fischer:2008wy} on 
the spectrum of light mesons. Another promising approach uses explicit representations 
of selected tensor structures of the quark-gluon vertex 
\cite{Fischer:2005en,Chang:2009zb,Chang:2010hb,Chang:2011ei}.
Most of these approaches have in common that they rely on techniques 
based on the two-particle-irreducible (2PI) representation of the 
effective action. In this language the interaction kernel is given as the
functional derivative of the quark self-energy with respect to the quark 
propagator as is detailed in Refs. \cite{McKay:1989rk,Munczek:1994zz}. 

In this work we use a similar idea. The difference is, though, that instead of 
employing a diagrammatic 
representation of the quark-gluon vertex, we use representations of the 
vertex that depend on the quark propagator explicitly and perform a systematic 
derivation of the corresponding Bethe-Salpeter kernel. Our approach is similar
in spirit but technically different from the one outlined in 
\cite{Chang:2009zb,Chang:2010hb,Chang:2011ei} and therefore serves as a 
complementary tool. In particular it has the advantage, that not only the 
mesons' masses but also their Bethe-Salpeter wave-function can be obtained.
This opens up the possibility for future studies of structural information such
as form factors and distribution amplitudes. In this respect, our approach 
improves upon the previous ones.

This paper is organized as follows. Section \ref{sec:General} contains basic 
definitions, central relations as well as the main theoretical ideas.
There we explain how we construct ladder and beyond-ladder kernels in general.
In Sections \ref{sec:2BCVertex}, \ref{sec:BCVertex} and \ref{sec:MunVertex}
we consider three specific vertex models, derive the corresponding interaction 
kernels and study the chiral properties of the different constructions. 
An implementation of these vertex models can be found in section \ref{sec:numerics} 
where our numerical results are presented. We conclude in section 
\ref{sec:conclusions}, followed by an appendix with technical details.

\section{Theoretical foundation}
\label{sec:General}
In this section we discuss the general principles that are 
at the heart of the techniques used in this work.
This will also serve to make some basic definitions 
and to introduce our notation.\\

\begin{figure}[t]
\resizebox{0.45\textwidth}{!}{%
 \includegraphics{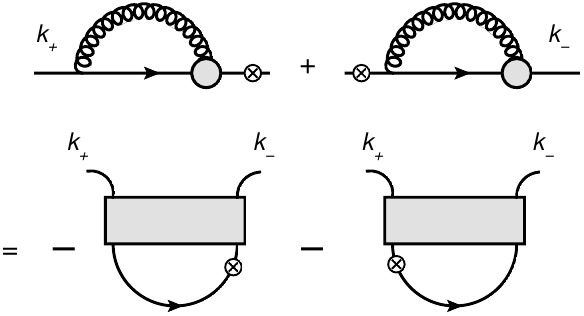}
}
\caption{A graphical representation of the AXWTI shown in Eq. (\ref{eqn:AXWTI2}). 
The grey dots are the dressed quark-gluon vertices, grey boxes denote the kernels 
and the crossed dots represent $\gamma_5$'s.}
\label{fig:AXWTI}       
\end{figure}

Our starting point is the definition of the quark
anti-quark interaction Kernel $K$ as the functional
derivative of the quark self-energy $\Sigma$ with
respect to the dressed quark propagator $S$
\begin{align}
  K^{cd}_{ab}(x,y,z,z^\prime) = \frac{\delta \Sigma^{cd}(x,y) }{\delta S^{ab}(z,z^\prime)},
  \label{eqn:KasFuntDerivative}
\end{align}
where $a, b, c$ and $d$ are Dirac indices and we work in coordinate space.
In a similar fashion, the quark self-energy is obtained from the 2PI effective
action. The technique given by Eq.~(\ref{eqn:KasFuntDerivative}) is often
called 'cutting' since in a graphical language it corresponds
to the cutting of a quark line. The 2PI
formalism allows for a closed representation of a truncated effective action
in terms of a loop expansion \cite{CJT,Berges:2004pu}. This has the advantage that
the validity of symmetries, such as chiral symmetries, can be checked on the level
of the effective action. 
The cutting procedure then generates equations that respect the consequences 
of the given symmetry. It has to be emphasized, however, that
cutting alone is not sufficient. The quark gluon vertex also needs to 
behave correctly under chiral transformations \cite{Munczek:1994zz}.
An appropriate tool to investigate the transformation properties of
the vertex in the momentum space representation is the 
axial-vector Ward Takahashi identity (AXWTI) which, if
fulfilled, guaranties a massless pion in the chiral
limit \cite{Maris:1997hd}.
In the chiral limit this identity reads
\begin{align}
  i P_\mu\Gamma_{5\mu}(P,k) = S^{-1}(k_+)\gamma_5 + \gamma_5 S^{-1}(k_-),
  \label{eqn:AXWTI}
\end{align}
where $\Gamma_{5\mu}(P,k)$ is the axial-vector vertex, depending on the total 
and relative quark momenta $P$ and $k$,
and $S^{-1}(k_\pm)$ the inverse quark propagator with $k_\pm =k\pm P$.
The axial-vector vertex has an exact representation via a Bethe-Salpeter
equation (BSE)
\begin{align}
  \Gamma_{5\mu}^{ab}(P,k)= -\int_q [S(q_+)\Gamma_{5\mu}(P,q)S(q_-)]^{cd}K_{cd}^{ab}(P,q,k),
  \label{eqn:AXVBSE}
\end{align}
where $K$ is the Fourier transform of the exact kernel defined
through Eq. (\ref{eqn:KasFuntDerivative}) and $\int_q = \int d^4 q/(2 \pi)^4$.
To proceed we have to define the quark self-energy in the exact form
\begin{align}
  \Sigma(k) = g^2 Z_{1F} C_F \int_q \gamma_\mu S(q) \Gamma_\nu(q,k) D_{\mu\nu}(q-k),
  \label{eqn:QuarkSelfEnergy}
\end{align}
with the Casimir $C_F = (N_c^2 - 1)/(2 N_c)$, the vertex renormalization factor $Z_{1F}$,
the gluon propagator $D_{\mu\nu}$ and the dressed quark-gluon vertex $\Gamma_\nu$
depending on the incoming and outgoing quark momenta.
The self-energy appears in the quark DSE
\begin{align}
  S^{-1}(k) = [S^0(k)]^{-1} + \Sigma(k).
  \label{eqn:QuarkDSE}
\end{align}
with inverse bare propagator $[S^0(k)]^{-1}=Z_2 (-i\sh{k}+m)$ including
the quark wave function renormalization factor $Z_2$.
The AXWTI from Eq. (\ref{eqn:AXWTI}) can be rewritten in the form 
\begin{align}
  &[\Sigma(k_+)\gamma_5 +\gamma_5\Sigma(k_-)]^{ab} =\notag\\
  &-\int_q [S(q_+)\gamma_5+\gamma_5S(q_-)]^{cd}
  K^{ab}_{cd}(P,q,k),
  \label{eqn:AXWTI2}
\end{align}
This representation is obtained upon inserting
the BSE Eq.~(\ref{eqn:AXVBSE}) in the AXWTI Eq.~(\ref{eqn:AXWTI}) and using
the Dyson-Schwinger equation of the quark Eq.~(\ref{eqn:QuarkDSE}).
A graphical representation of the resulting expression can be found in
Fig.~\ref{fig:AXWTI}.

Before we apply the derivative of Eq. (\ref{eqn:KasFuntDerivative})
to a given vertex representation we make a short mathematical
detour. The space of Euclidean Dirac-matrices with Clifford algebra 
$\{\gamma_\mu,\gamma_\mu\}=2\delta_{\mu\nu}$ is spanned by
the 16 dimensional basis \linebreak
${T_i}=\{\gamma_{\mu},  \OperatorOne ,\gamma_5, \gamma_5\gamma_\mu,\sigma_{\mu\nu}\}$
with $\sigma_{\mu\nu}=i/2[\gamma_\mu,\gamma_\nu]$. The elements 
obey $T_iT_i=\OperatorOne$ (no summation of indices) 
and $1/4\,\,\tr[T_iT_j]=\delta_{ij}$. Thus in general, the fully 
dressed quark propagator and inverse propagator can be represented by 
\begin{xalignat}{2}
  S(p) &= \sum_{i=1}^{16}T_i \tau_i(p^2)\,, & 
  S^{-1}(p)& = \sum_{i=1}^{16}T_i \mathcal{A}_i(p^2)\,,
  \label{eqn:QuarkDecomposition}
\end{xalignat}
where the quark dressings $\tau$ and $\mathcal{A}$ depend on the quadratic 
momentum only. The physical quark propagator has the structure 
$S(p) = i \sh{p}\sigma_V(p^2) + \sigma_S(p^2)$ but
in the process of taking the derivative the representation
of Eq.~(\ref{eqn:QuarkDecomposition}) is necessary for reasons of
completeness. In particular we wish to maintain
\begin{align}
 \delta_{ac}\,\delta_{bd}\,  \delta^{(4)}(p-q) &\stackrel{!}{=} \frac{\delta S^{ab}(p)}{\delta S^{cd}(q)} \notag\\
&= \sum_{i=1}^{16}\frac{\delta \tau_i(q)}{\delta S^{cd}(q)} \frac{\delta}{\delta \tau_i(q)}
 \sum_{j=1}^{16} T_j^{ab}\tau_j(p)\notag\\
 &=\sum_{i=1}^{16}\frac{1}{4}T_i^{dc}T_i^{ab} \delta^{(4)}(p-q),
  \label{eqn:DerivativeIdentity}
\end{align}
which, as a completeness relation, can only be valid with the full basis.
We used $\delta \tau_i/\delta S^{cd} =1/4[T_i^{dc}]^{-1}=1/4T_i^{dc}$
and $\delta \tau_j(p)/\delta \tau_i(q) = \delta_{ij}\delta^{(4)}(p-q)$.

The functional derivative onto the quark self-energy Eq.~(\ref{eqn:QuarkSelfEnergy})
acts on the quark itself, the vertex and the gluon. For simplicity, in the 
following we disregard derivatives of the gluon propagator. Since the gluon 
depends on the quark only implicitly via closed loops, contributions 
from derivatives with respect to the quark only show up in kernels of flavor-singlet mesons. 
The following discussion is therefore directly applicable only in non-flavor-singlet
channels but can be easily generalized to include also the flavor-singlet case. 

Although the cutting rule is probably best defined in coordinate space,
let us first work in momentum space. On the one hand, this serves illustrational 
purposes, on the other hand this is necessary for vertex models such as the
Ball-Chiu construction \cite{Ball:1980ay}, which are derived in momentum space. 
Later on we will demonstrate that the cutting procedure is much simpler in 
coordinate space and elaborate on a vertex construction (the Munczek vertex 
\cite{Munczek:1994zz}) that has a corresponding representation. 
Cutting the quark propagator, we obtain the {\it modified} ladder-like contributions
(called type $I$ in the following)
\begin{align}
  \left.\frac{\delta \Sigma^{ab}}{\delta S^{cd}}\right|_{I}=
  \gamma_\mu^{ac}D_{\mu\nu}\Gamma_\mu^{db}\,.
  \label{eqn:LadderKernel}
\end{align}
This corresponds to a nonperturbative one-gluon exchange. However, in contrast
to the usual ladder kernels one of the quark-gluon vertices is dressed. 

For brevity, the kinematic dependences in Eq.~(\ref{eqn:LadderKernel}) are suppressed. 
On a diagrammatic basis the correct 
kinematics are easily determined. Yet on a strict mathematical basis the 
quark is an arbitrary function in the 2PI formalism. Translational invariance 
cannot be assumed before relaxing the quark to the physical point. Thus in general
one has to allow the quark and the self-energy to depend on different ingoing
and outgoing momenta. It will be seen below in sections \ref{sec:2BCVertex} and
\ref{sec:BCVertex}, that this introduces some complications in the kinematical 
dependencies of our kernels. These problems are easily overcome when working in
coordinate space as will be shown in section \ref{sec:MunVertex}.

Note that with a pure kernel of type $I$ the AXWTI from Eq.~(\ref{eqn:AXWTI2}) is 
fulfilled in the limit $P\rightarrow0$ only if \linebreak $\{\Gamma_\mu,\gamma_5\}=0$. This is 
because the terms on the right side of the AXWTI assume the form of self-energies
for type $I$ kernels. The $\gamma_5$ has to be moved past the vertices, however.
This is trivial for the bare vertex, but non-trivial for more elaborate vertex 
constructions, pointing towards the necessary appearance of a further type of contributions.

Indeed, a second type of contributions to the interaction kernel contain the variation of 
the quark-gluon vertex
\begin{align}
  \left.\frac{\delta \Sigma^{ab}}{\delta S^{cd}}\right|_{II}=
  \int_q [\gamma_\mu S(q)]^{aa^\prime} \frac{\delta \Gamma_\mu^{a^\prime b}(q,p)}{\delta S^{cd}(s)}D_{\mu\nu}(p-q),
  \label{eqn:BeyondLadderKernel}
\end{align}
which is referred to as type $II$ contribution.
In our notation the variation of the vertex can be decomposed as
\begin{align}
   \frac{\delta \Gamma_\mu^{ab}}{\delta S^{cd}} &=
   \sum_{i} \frac{1}{4}T^{dc}_i \frac{\delta\Gamma^{ab}_\mu}{\delta \tau_i}
  \label{eqn:VertexCuttingDecomposition}
\end{align}
where the Dirac indices $\{c,d\}$ are the ones connecting to the incoming 
quarks. Thus the appearance of certain Dirac structures in the interaction 
kernel is dictated by whether a corresponding functional variation of the 
vertex evaluates to zero or not. We will come back to this in the following 
sections where different vertex representations are considered.

The main observables that we will study to underline our theoretical
considerations and test the approach are masses of light mesons in the
(pseudo-)scalar and (axial-) vector channels. Their generic Bethe-Salpeter 
equation for the meson amplitude $\Gamma_M^{(\nu)}$ is given by
\begin{align}
  &\left[\Gamma_M^{(\nu)}\right]^{ab}(P,k)=\notag \\
  &-\int_q [S(q_+)\Gamma_M^{(\nu)}(P,q)S(q_-)]^{cd}K_{cd}^{ab}(P,q,k),
  \label{eqn:PionBSE}
\end{align}
with kernel $K$ and the total momentum satisfies $P^2=-m_M^2$ with $m_M$ the
mass of the meson in question. For pseudoscalar mesons, like the pion, the 
amplitude has the decomposition 
\begin{align}
  &\Gamma_\pi(P,k)=\notag\\
  &\gamma_5\left[ E(P,k) + i \Sh{P} F(P,k)
  + i \sh{k} G(P,k) - [\Sh{P},\sh{k}] H(P,k)  \right].
  \label{eqn:PionBSA}
\end{align}
Similar decompositions for the other mesons are given e.g. in Ref.~\cite{Alkofer:2002bp}.
Furthermore we quote here the Gell-Mann--Oakes--Renner relation (GMOR) \cite{GellMann:1968rz} 
\begin{align}
  f_\pi^2 m_\pi^2 = \langle\bar{\psi}\psi\rangle_{\mu} m(\mu),
  \label{eqn:GMOR}
\end{align}
which will be a tool to test the chiral properties in our numerical treatment in section
\ref{sec:numerics}. Here $f_\pi$ is the pion decay constant, $\langle \bar{\psi}\psi\rangle_{\mu}$ 
the chiral condensate and $m(\mu)$ the running quark mass at renormalization point $\mu$.

\section{Constructing the kernel}

In the following we show explicitly, how our formalism serves to construct the kernel,
once a representation of the quark-gluon vertex in terms of the quark dressing functions
is known. For the longitudinal part of the vertex such a representation can be derived
(approximately) from its Slavnov-Taylor identity \cite{Eichten:1974et}.
It reads
\begin{eqnarray}
p_3^\mu \Gamma(p_1,p_2) &=& G(p_3^2) \times\\
&& \times \left[H(p_1,p_2)S^{-1}(p_2) - S^{-1}(p_1) H(p_1,p_2)\right] \nonumber
\end{eqnarray}
in terms of the inverse quark propagator $S^{-1}$, the ghost dressing function 
$G$ and a ghost-quark scattering kernel $H$. The momenta $p_1,p_2$ correspond to the 
quark legs of the vertex, whereas $p_3=p_2-p_1$ denotes the momentum from the gluon leg. Assuming
that $H(p_1,p_2)$ can be approximated by a function $\tilde{H}(p_3^2)$ depending on the
gluon momentum only, the STI can be converted into a Ward-Takahashi identity with an extra
factor $G\,\tilde{H}$ on the right hand side. It is then solved by the Ball-Chiu construction 
\cite{Ball:1980ay} supplemented with the product $G\,\tilde{H}$
\begin{eqnarray}
  \Gamma_\mu^{BC}(p_1,p_2) &=& G(p_3^2) \tilde{H}(p_3^2)\left[ \gamma_\mu \frac{A(p^2_1)+A(p^2_2)}{2} \right.
  \label{BC}\\
  &&\left.\hspace*{-1cm}+ 2\sh{k}k_\mu \frac{A(p^2_1)-A(p^2_2)}{p_1^2-p_2^2}
          + i 2 p_\mu \frac{B(p^2_1)-B(p^2_2)}{p_1^2-p_2^2}\right]\,,\nonumber
  \label{eqn:3rdBCVertex}
\end{eqnarray}
where $k=(p_1+p_2)/2$ and vector dressing $A$ and scalar dressing $B$ of the inverse
quark propagator 
\begin{equation}
S^{-1}(p)=-i\sh{p}A(p^2)+ \OperatorOne B(p^2)\,.
\end{equation} 
Within the quark-DSE the functions $G(p_3^2) \tilde{H}(p_3^2)$ can then be combined 
with the gluon propagator into an effective gluon (cf. Appendix \ref{app:gluon})
and the vertex has an Abelian structure.
As a result one has a representation of the vertex in terms of the quark dressing functions.
In general, this construction can be supplemented by transverse terms that are not restricted
by the STI/WTI and can be either modeled or extracted from explicit solutions of (approximations
of) the vertex-DSE. However, for the purpose of this work we restrict ourselves to the
Ball-Chiu part of the vertex since it serves nicely to illustrate the merits of our formalism.

In the following we will first treat the first two terms of this vertex ('2BC vertex model'),
then deal with the third term in addition ('Ball Chiu vertex model') and finally work with a 
different solution of the WTI (the 'Munczek vertex model') that is suited to explore the cutting
procedure in coordinate space.

\subsection{The 2BC Vertex model}
\label{sec:2BCVertex}
Here we consider the first two terms of the parts of the Ball-Chiu vertex 
Eq.~(\ref{BC}) with tensor structures $\gamma_\mu$ and $\sh{k}$. Note that these 
structures correspond to four different structures in the notation of 
Eq.~(\ref{eqn:QuarkDecomposition}).
In order to carry out the cutting for type $II$ kernels along the lines of
Eqs.~(\ref{eqn:BeyondLadderKernel}) and (\ref{eqn:VertexCuttingDecomposition})
we therefore write
\begin{xalignat}{3}
    \gamma_\mu A&\rightarrow \gamma_\mu A_\mu 
    & \sh{k} A   &\rightarrow \sum_{\alpha}k_\alpha\gamma_\alpha A_\alpha
    & \mu,\alpha&\in\{1,2,3,4\},
  \label{eqn:2BCunphysical}
\end{xalignat}
where no summation over the index $\mu$ is performed. The functions
obey $A_{\mu}=\mathcal{A}_{\mu}/(-ip_\mu)$, i.e. they represent the subset
of the $\mathcal{A}$ functions, defined in section \ref{sec:General}, 
that correspond to the $\gamma_\mu$ structures of the Dirac algebra. Via 
Eq.~(\ref{BC}) the $A_\mu$ functions are explicitly given in terms of the 
$\tau$-dressings of the quark in the notation of Eq.~(\ref{eqn:QuarkDecomposition}). 

The vertex model Eq.~(\ref{BC}) is defined on the physical point of Dirac space
and we call the four contributing basis structures $\gamma_i$ the physical directions
in Dirac-space. However, the actual cutting procedure Eq.~(\ref{eqn:KasFuntDerivative}) 
has to be performed in all directions of Dirac-space, i.e. also in the unphysical 
ones. In analogy to ordinary functions a functional that is zero at
a given point may nevertheless have a non-vanishing functional derivative. 
Thus the cutting-procedure may very well pick up contributions from the
unphysical directions. In order to completely specify a vertex model {\it and}
the corresponding  Bethe-Salpeter kernel it is therefore not sufficient to define
the model on the physical point, but we need additional information on its
behavior in the unphysical directions of Dirac-space.

\begin{figure}[t]
\includegraphics[width=\columnwidth]{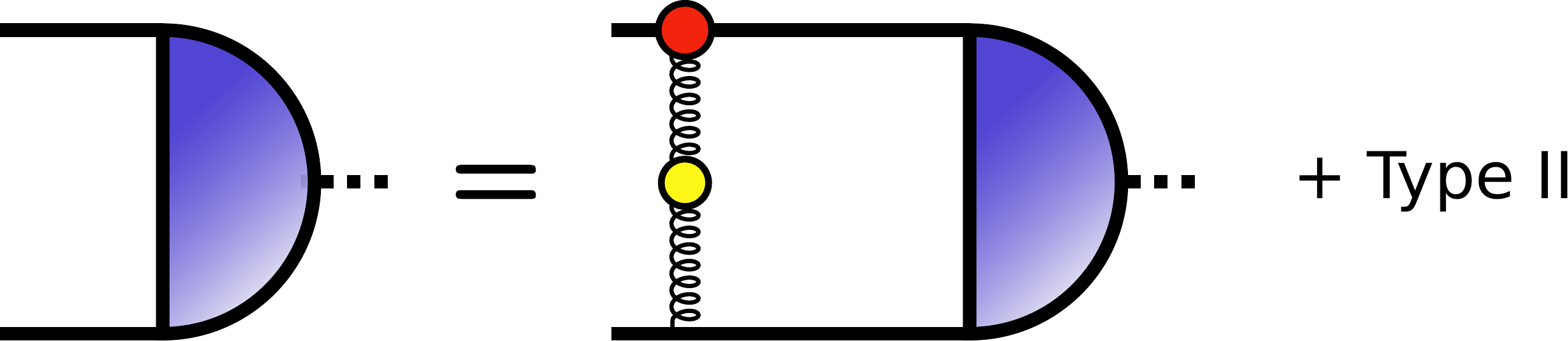} 
\caption{Bethe-Salpeter equation for mesons including kernel contributions of type I 
(dressed one-gluon exchanged) and type II (see text).}
\label{fig:BSE}       
\end{figure}

This is irrelevant for the type I contribution of the kernel obtained from cutting the
quark line in the quark-DSE. The resulting expression is afterwards set to the
physical point and represents the modified dressed one-gluon exchange shown in 
Fig.~\ref{fig:BSE}.
The situation is different, however, for the type II contributions involving the
functional derivative of the vertex.

Let us assume for the moment that our 2BC vertex model away from the physical point still 
has the reduced functional dependence $\Gamma_\mu^{2BC}[A_\mu[\tau_{1\dots 4}]]$ corresponding
to $T_{1\dots4}=\gamma_{1\dots4}$ and the unphysical directions in Dirac space are identical 
to zero. This then yields $\delta \Gamma_\mu/\delta \tau_i=0, \forall i>4$ such that the 
external legs of the kernel that will connect to the internal quark lines in the Bethe-Salpeter
equation have a restricted tensor structure. In this case, type II contributions to the kernel
will appear, but due to their restricted tensor structure they contribute neither
to the AXWTI nor to the Bethe-Salpeter equation for pseudoscalar (and axialvector) mesons. 
This is because of the $\gamma_5$ contained on the right hand side of the AXWTI Eq.~(\ref{eqn:AXWTI2})
and in the meson amplitudes of the Bethe-Salpeter equation (\ref{eqn:PionBSE}) which lead
to zero traces. The explicit form of these type II contributions, relevant for scalar and 
vector mesons, is discussed in appendix \ref{app:Kernels}. 

In the AXWTI we thus have to consider only the modified ladder type contributions.
Both vertex structures in the first two terms of Eq.~(\ref{BC}) anti-commute
with $\gamma_5$ so that for this particular vertex model the AXWTI Eq.~(\ref{eqn:AXWTI2}) 
is fulfilled in the limit $P\rightarrow 0$. For the pseudoscalar bound states the 
modified ladder contributions are all that remains and lead to a massless pion in
the chiral limit. This finding will be confirmed by our numerical results in 
section \ref{sec:numerics}.

Note, however, that the AXWTI is only satisfied in the limit $P\rightarrow 0$. This is 
because the left and right side of the equation, although similar on the diagrammatic 
level, need a momentum shift to be absolutely identical. This momentum shift becomes 
impossible due to the momentum dependence of the first two terms in the vertex (\ref{BC}). In the 
limit $P\rightarrow 0$, however, the diagrams become equal. For physical pions, a simple
vertex model such as the 2BC-vertex cannot be the full story and corrections from
unphysical directions in Dirac space are necessary. In principle, the requirements of
chiral symmetry via the AXWTI allow for a systematic procedure to construct such
extensions thus completing a given vertex model. We will perform this exercise in
the next section. Allowing the functions $A_\mu$ to depend on $\tau_{1\dots16}$, 
non-vanishing contributions of type $II$ in the AXWTI and the pseudoscalar BSEs are 
generated which can be used to restore the requirements of chiral symmetry also away 
from the chiral limit. This emphasizes again that a truncation is not uniquely fixed 
by the vertex model on the physical point. 

\subsection{The Ball-Chiu vertex model}
\label{sec:BCVertex}
In addition to the first two terms of the Ball-Chiu type vertex in Eq.~(\ref{BC})
we will also consider the third term, which is proportional to the scalar basis element
$T_5=\OperatorOne$ in Dirac space. Since this term does not anti-commute with 
$\gamma_5$ it cannot fulfill the AXWTI (\ref{eqn:AXWTI2}) on the level of a pure  
type $I$ ladder kernel. This can, however, be cured by allowing for type $II$ 
contributions to the kernel that couple to the pseudo scalar channel. In order
to generate these in a systematic way, we allow the vertex to depend on unphysical 
components that will be set to zero in the end, but will contribute during the cutting 
procedure. Therefore, we write the quark generically as
\begin{align}
  S(p) &= \left(i\sum^4_{j=1} \sigma_j\, p_j\, \gamma_j\right) + \sigma_S\, \OperatorOne
  + \sigma_5\, \gamma_5\notag\\
  S^{-1}(p)& = \left( -i\sum^4_{j=1}A_j\, p_j\,\gamma_j \right) + B \,\OperatorOne
  + C\, \gamma_5\,,
  \label{eqn:unphysQuark}
\end{align}
with $\sigma_5=0$, $C=0$, $\sigma_j=\sigma_V$ and $A_i=A$ on the physical point. The 
functions obey $\sigma_{j}=\tau_{j}/ip_j, \forall i\leq 4$, $\sigma_S=\tau_5$ and $\sigma_5=\tau_6$.
The reason why we need only six instead of the full sixteen tensor structures in Dirac space 
is that we will assume a certain functional dependence of the vertex on the quark dressings 
as in the preceding section. Our vertex, called {\it $ABC$-vertex} from now on, reads
\begin{align}
  \Gamma_\mu^{ABC} = \Gamma_\mu^{BC} + i2\, \gamma_5 \,k_\mu\frac{C(k^2_+)-C(k^2_-)}{k_+^2-k_-^2}.
  \label{eqn:ABCVertex}
\end{align}
This vertex corresponds to the Ball-Chiu construction for a quark with $C\neq 0\neq \sigma_5$ 
as given in Eq. (\ref{eqn:unphysQuark}). Thus we assume that the vertex does depend only on 
the quark dressings $A$, $B$ and $C$, limiting the possible structures in the type $II$ part
of the kernel. We furthermore generalize the $A$ function as discussed in Eq.~(\ref{eqn:2BCunphysical}). 
This fully determines the Ball-Chiu type of vertex construction for a quark of the form shown
in equation (\ref{eqn:unphysQuark}).

Now we have laid the basis to explicitly derive
the type $II$ kernel for the vertex of Eq. (\ref{eqn:ABCVertex}).
The complete set of these kernels is treated in appendix \ref{app:Kernels}.
It turns out, however, that the application of Eq. (\ref{eqn:VertexCuttingDecomposition})
 generates only one single type $II$ kernel that contributes to the AXWTI
and the pion BSE after relaxing all dressings to the physical case. Only the derivative
with respect to $\sigma_5$, being accompanied by the $\gamma_5^{dc}$ structure (see Eq. (\ref{eqn:VertexCuttingDecomposition})), will give a non-zero contribution upon tracing with the 
additional $\gamma_5$ as present in the AXWTI (\ref{eqn:AXWTI2}) and the pion BSA 
(\ref{eqn:PionBSA}). 

The relevant piece of Eq. (\ref{eqn:VertexCuttingDecomposition}) evaluates to
\begin{equation}
   \frac{1}{4}\gamma_5^{dc}\frac{\delta \Gamma^{ab}_\nu(l,k)}{\delta \sigma_5(q)} =
  \frac{1}{4}\frac{i  (l+k)_\nu}{l^2-k^2}\left[ \frac{\delta C(l^2)}{\delta \sigma_5(q)}
  -   \frac{\delta C(k^2)}{\delta \sigma_5(q)}\right]\,  \gamma_5^{dc}\,\gamma_5^{ab},
  \label{eqn:BRLKernelC5}
\end{equation}
with
\begin{align}
% & \frac{1}{4} \left.\frac{\delta C(p)}{\delta \sigma_5(q)}\right|_{phys} [\gamma_5]^{ab}\,\, [\gamma_5]^{cd}\notag\\
  \left.\frac{\delta C(l^2)}{\delta \sigma_5(q)}\right|_{phys}&=-\frac{1}{4}
  \frac{1}{\sigma_V^2(l) l^2 +\sigma_S^2(l) }\, \delta^{(4)}(l-q).
  \label{eqn:BRLKernel}
\end{align}
The corresponding kernel is then generated by insertion into
Eq. (\ref{eqn:BeyondLadderKernel}). The resulting expression is 
provided in  appendix \ref{app:AXWTI3BC}  where we also prove that the
AXWTI is fulfilled in the limit of vanishing
total momentum.\\

We would like to emphasize that we make a non-trivial
observation here. We nicely see how 
type $I$ and type $II$ contributions cancel each other
exactly in the AXWTI as is shown explicitly in appendix \ref{app:AXWTI3BC}.
This gives deep insight into the way chiral symmetry is at work
in beyond rainbow-ladder truncations in general. In fact it is no
coincidence that the generalized Ball-Chiu vertex from equation (\ref{eqn:ABCVertex})
has the correct behavior. Following the arguments of Ref. \cite{Munczek:1994zz}
a quark-gluon vertex model that transforms under local chiral transformations
as an inverse quark should leave chiral symmetry intact in every possible
relation derived from the 2PI effective action. A vertex that fulfills
the vector WTI, as the BC vertex does, is thus at least a very good
candidate for a vertex model. We show here, how these formal arguments are 
realized explicitly in a Bethe-Salpeter interaction kernel.

There is, however, an additional subtle point here. The momentum space representation of
Eq (\ref{eqn:KasFuntDerivative}), if written as $K = \delta \Sigma(p)/\delta S(l)$,
depends only on two momenta, $p$ and $l$. As a four-point function $K$ should depend
on three independent momenta in general $K(P,p,l)$ as is the case for the type $I$
interaction $K_I(P,p,l) = \gamma_\mu D_{\mu\nu}(p-l)\Gamma_\nu(l,p_+)$.
As argued above, from our cutting procedure this is not plain obvious, instead
we complete the kinematic dependence on the diagrammatic level. For the type $II$
kernels this is, however, not so simple since these have no representation
as Feynman-diagrams. We choose the kinematics such that the potential
dangerous singular structure of Eq. (\ref{eqn:BRLKernelC5}) stays
harmless. This is also detailed in appendix \ref{app:AXWTI3BC}.

\subsection{The Munczek vertex model}
\label{sec:MunVertex}
Finally we treat a vertex model that has been formulated in coordinate space. We
will see, that this choice leads to unique kinematics in the derived kernel and
provides for a simple and elegant kernel. The vertex ansatz has been given by Munczek 
in Ref.\cite{Munczek:1994zz} and reads in coordinate space:

\begin{align}
 \Gamma_\mu(z;x,y) & = i S^{-1}(x,y) \times\label{eqn:MunczekVertex} \\
   & \times \int \frac{d^4 q}{(2\pi)^4} \left[  e^{iq\cdot(z-y)} - e^{iq\cdot(z-x)} \right]\frac{x_\mu - y_\mu}{q\cdot(x - y)}.\nonumber
  \end{align}

Because of the unusual form of this vertex, we repeat a few arguments for this 
particular choice of vertex, given in  \cite{Munczek:1994zz}:\\
This vertex transforms under local chiral transformations in the following way:
\begin{equation}
  \Gamma_\nu (z;x,y) \rightarrow e^{-i\gamma_5\tau^l\theta_l(x)}\Gamma_\nu(z;x,y)e^{-i\gamma_5\tau^l\theta_l(y)}
 \label{eqn:VertexTrafo}
\end{equation}
similar to the inverse quark
\begin{equation}
  S^{-1} (x,y) \rightarrow e^{-i\gamma_5\tau^l\theta_l(x)}S^{-1}(x,y)e^{-i\gamma_5\tau^l\theta_l(y)}.
 \label{eqn:QuarkTrafo}
\end{equation}

This ensures that the 2PI effective action is invariant under a local chiral 
transformation which is necessary for the pion to be a Goldstone boson.
As in the Ball-Chiu case, this vertex ansatz is free of kinematical singularities
and compatible with the vector Ward identity (WTI) in coordinate space:
\begin{equation}
 \frac{\partial}{\partial z_\mu}\Gamma_\mu(z;x,y) = i\left[ \delta(y-z) - \delta(x-z)  \right]S^{-1}(x,y).
\end{equation}
From a technical point of view, the biggest advantage of this vertex, in 
comparison to the Ball-Chiu one, is the fact, that a representation in 
coordinate space is available. All the problems of the ambiguous momentum 
routing, that plagued the cutting procedure for the Ball-Chiu vertex are 
resolved when applying the cutting procedure to the selfenergy 
in coordinate space. After the cutting, a transformation back to momentum 
space is possible and yields a closed expression for the interaction kernel 
and the vertex. For more technical details we refer to appendix \ref{app:MunModel},
presenting here only the results.

With the definition
\begin{equation}
  \left[\hat{S}^{-1}\right]^\mu := \frac{\partial}{\partial k^\mu} S^{-1}(k)\Big|_{k=k_r+\alpha (k_l - k_r)}
\end{equation}
the vertex rads  
\begin{equation}
  \Gamma^{\mu}(k_l,k_r) =i\int\limits^1\limits_0\left[\hat{S}^{-1}\right]^\mu\, d\alpha.
  \label{eq:VertexMunMomSpace}
\end{equation}
Here the momentum $k_l$ specifies the incoming left momenta and $k_r$ the outgoing 
right momenta. They are connected via  $k=k_l-k_r$ where $k$ is the outgoing
gluon momentum.
For the kernel we introduce some shorthand notations:
\begin{align}
 \left[\hat{\Gamma}_\pi\right]^\nu &:=  \frac{\partial}{\partial p^\nu} \Gamma_\pi(p;P)\Big|_{p=\tilde{p}+\alpha (\tilde{p}-p)}\\
\tilde{\Gamma_\pi}  &:= S(p_+)\Gamma_\pi(P,p)S(p_-),,
\label{eqn:WaveFunction}
\end{align}
with $P$ the total and $p$ the relative momenta of the two-body bound state
and $p_\pm = p \pm P/2$.
The resulting kernel can be written down in a closed form as linear operator.
 Inserted into
the right hand side of the Bethe-Salpeter equation we obtain 
\begin{align}
   &\int [K^{II}\tilde{\Gamma_\pi}]_{a b}= \nonumber\\  
  & \frac{i}{2}\int\limits_{\substack{\tilde{p}\\}}d^4\tilde{p}\int\limits_0\limits^1d\alpha\left[\hat{\Gamma}_\pi\right]^\nu_{b^\prime b}S_{a^\prime b^\prime}(\tilde{p}_- )\gamma_{a a^\prime}^\mu  D^{\mu\nu}(\tilde{p}-p) \nonumber\\
  & \phantom{\frac{i}{2}\int\limits_{\substack{\tilde{p}\\}}d^4\tilde{p}\int\limits_0\limits^1d\alpha}+\left[\hat{\Gamma}_\pi\right]^\nu_{a a^\prime} S_{a^\prime b^\prime}(\tilde{p}_+ )\gamma_{b^\prime b}^\mu D^{\mu\nu}(\tilde{p}-p).
 \label{eqn:Munczekvertex} 
\end{align}
This expression is already symmetrized as explained in appendix \ref{app:MunModel}. The latin indices represent the Dirac matrix 
indices. All other additional factors as color etc. are suppressed. It is interesting 
to see that the selfenergy of the quark, Eq. (\ref{eqn:QuarkSelfEnergy}) with 
Eq. (\ref{eq:VertexMunMomSpace}) as vertex, and the type II contribution to 
the BSE, Eq. (\ref{eqn:Munczekvertex}),
have the same structure. The only difference, modulo momentum dependence, is the 
replacement of $ \left[\hat{S}^{-1}\right]^\mu$ with $\left[\hat{\Gamma}_\pi\right]^\mu$. 
Upon inserting the kernel in Eq.~(\ref{eqn:AXWTI2})
and working out the details it can be seen, that the AXWTI is fulfilled and that 
the structural similarity plays an important role in doing so. 

These findings can be summarized in the following way:
\begin{itemize}
 \item In order to meet the transfomation property of Eq.~(\ref{eqn:VertexTrafo})
       a vertex model is chosen that depends linearly on $S^{-1}$ with the transformation properties
       of Eq.~(\ref{eqn:QuarkTrafo}).
 \item The additional terms on the rhs of Eq.~(\ref{eqn:AXWTI2}), stemming from the cutting procedure
       have the same structure as a quark selfenergy \emph{because} the vertex is linear dependent on
       $S^{-1}$.
 \item This additional terms that look like quark selfenergies cancel other terms on the rhs of Eq. (\ref{eqn:AXWTI2}).
       This happens in a similar fashion as for the ABC vertex, cf. appendix \ref{app:AXWTI3BC}.  
\end{itemize}

As shown in Ref.~\cite{Munczek:1994zz} this comes with no surprise: If the 
vertex transforms in the proper way, the determination of the kernel via 
cutting of the quark selfenergy yields a interaction that preserves the AXWTI. 
The linearity on $S^{-1}$ is not necessary, but one has to work much
harder to preserve the correct transformation behavior if the vertex is nonlinear in $S^{-1}$.
In the result section we check the GMOR explicitly for this particular choice of vertex.

We make a last comment regarding the similarity between the Munczek vertex and the BC vertex. 
Despite the unusual form of the vertex in Eq.~(\ref{eqn:Munczekvertex}), this vertex has a 
striking resemblance with the Ball-Chiu vertex in momentum  space. This can be seen by 
carrying out the derivative in Eq.~(\ref{eq:VertexMunMomSpace})
explicitly
\begin{align}
  \Gamma^{\mu}(p_l,p_r) = \int\limits^1\limits_0 d&\alpha\big[ 2p^\mu A^\prime(p^2)\sh{p} \\
  & \gamma^\mu A(p^2) + i2p^\mu B^\prime(p^2) \big]_{p=p_r+\alpha (p_l - p_r)}.\nonumber
\end{align}
Where the BC vertex has terms that look like finite differences, the Munczek vertex 
has derivatives smeared by the $\alpha$ integral.  

\section{Numerical results}
\label{sec:numerics}

For our numerical analysis the quark DSE Eq. (\ref{eqn:QuarkDSE}) was solved for 
complex momenta following a contour method, described in Ref.~\cite{Krassnigg:2009gd}.
The BSE is solved as an eigenvalue problem with standard numerical methods. 
\begin{figure}[t]
\resizebox{0.49\textwidth}{!}{\includegraphics{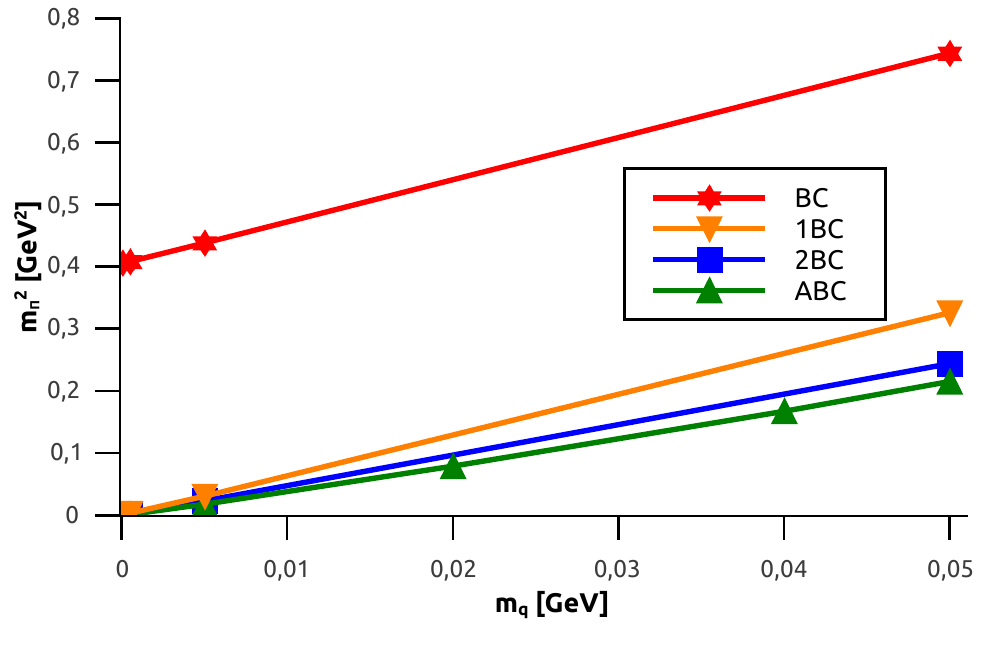}}
\caption{This figure depicts the dependence of the squared pion mass on the quark mass 
in the BC-vertex 
models. Here '1BC' corresponds to a vertex, where only the first term of the Ball-Chiu
vertex has been taken into account, '2BC' to the vertex treated in section \ref{sec:2BCVertex},
'BC' is the physical Ball-Chiu vertex dealt with in section \ref{sec:BCVertex} and 
'ABC' is its completion with unphysical directions before cutting, Eq.~(\ref{eqn:ABCVertex}).
The quark mass $m_q$ is evaluated at a renormalization point of $\mu = 19$ GeV.}
\label{fig:BC_results}       
\end{figure}

We solved the BSE for the different Ball-Chiu vertex models described in the sections 
before. Our first main result is shown in Fig. \ref{fig:BC_results}.
For the 1BC and 2BC vertices, the cutting of the vertex yields no additional 
contribution to the kernel of the pion, so that the kernel is purely of type $I$.
As argued before, the 1BC and 2BC vertex models have only vector contributions that 
are proportional to $\gamma^\mu$ and thus cause no problems in the AXWTI. As one can 
see in Fig.~\ref{fig:BC_results} the GMOR-relation Eq.~(\ref{eqn:GMOR}) is satisfied:
the squared pion mass scales linearly with the quark bare mass and goes through the 
origin.

The results for the full BC vertex are much more intricate. As described above, setting 
the BC vertex to its physical form before the cutting procedure yields no contribution 
of type $II$ to the pion BSE. Since the scalar parts of the vertex proportional to 
$\mathbbm{1}$ spoil the AXWTI, the resulting equations are not in accord with the 
requirements of chiral symmetry. This directly translates to a severe violation of the 
GMOR with a heavy pion of $m_\pi = 637\,$ MeV in the chiral limit. Adding 'unphysical' directions 
to the vertex before cutting, however, solves the problem as elaborated in section 
\ref{sec:BCVertex}. With the resulting ABC-vertex the GMOR-relation is satisfied,
and the corresponding curve in Fig.~\ref{fig:BC_results} again describes a Goldstone boson.

\begin{figure}[t]
\resizebox{0.49\textwidth}{!}{\includegraphics{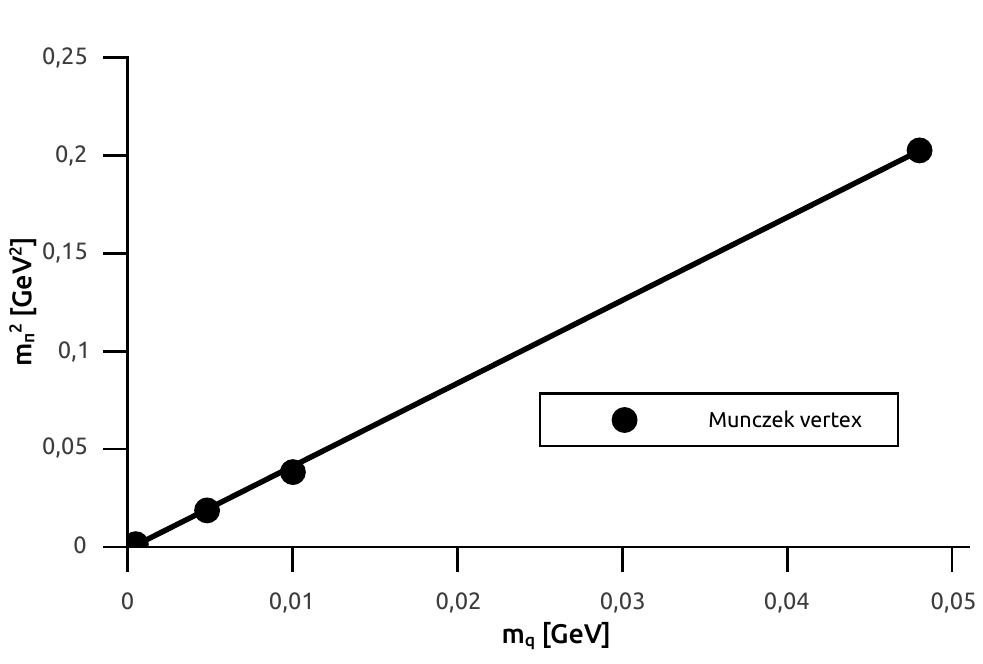}}
\caption{This figure depicts the dependence of the squared pion mass on the quark mass 
in the Munczek-vertex model.}
\label{fig:Mun_results1}       
\end{figure}

Our results for the Munczek vertex model from section \ref{sec:MunVertex} are displayed 
in Fig.~\ref{fig:Mun_results1}. Again we find that the pion becomes a Goldstone boson in 
the chiral limit. The main difference as compared to the Ball-Chiu vertex is that we did
not have to add terms along unphysical directions. This nicely underlines the main message 
of Ref.~\cite{Munczek:1994zz}: having the right chiral transformation property already
on the level of the vertex representation ensures a massless pion in the chiral limit,
provided the kernel is properly constructed. The Munczek vertex preserves all symmetries 
by design and produces the correct type $II$ kernel contributions for the pion automatically.

It is also interesting to discuss the physical implications of such a type of vertex.
The Munczek vertex possesses a structure that is proportional to $\mathbbm{1}$ and
the derivative of the scalar quark dressing function. Such a structure is not present 
in the chirally symmetric theory and thus represents an important addition to the 
structure of the vertex that is mainly generated by the dynamical effects of chiral
symmetry breaking. This structure is also not present in a usual ladder approximation 
with a vertex proportional to $\gamma^\mu$. In the Munczek vertex, this term plays a 
similar role than the corresponding scalar contribution to the Ball-Chiu vertex.
In Ref.~\cite{Chang:2009zb} this term has been interpreted as being responsible
for a dramatic increase in the mass splitting between the scalar and the pseudoscalar 
ground state and was interpreted as a repulsive spin-orbit force. 

\begin{figure}[t]
\includegraphics[width=\columnwidth]{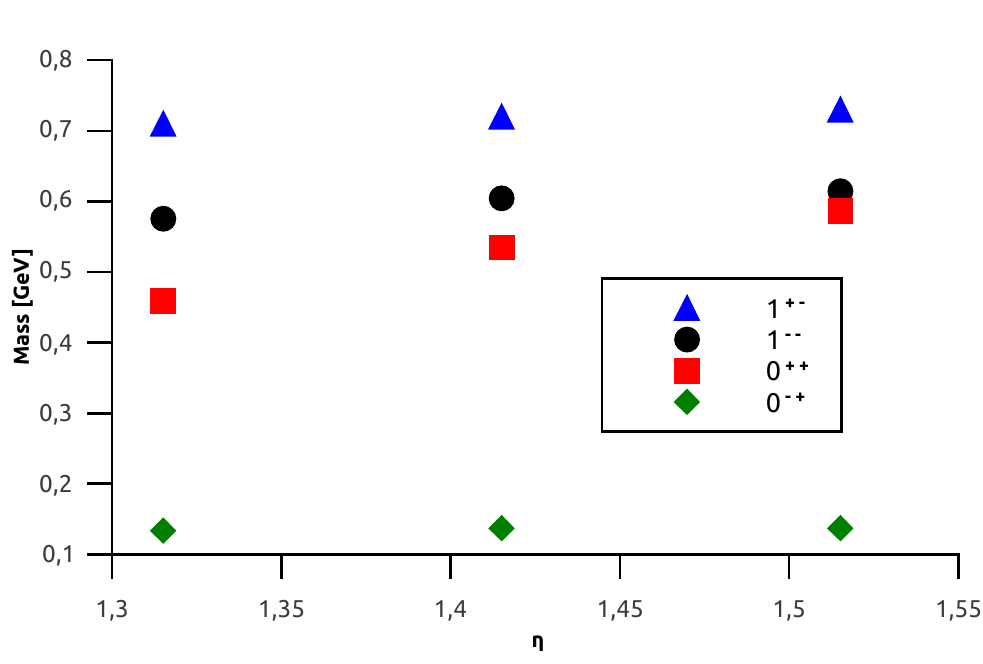}\hfill  
\includegraphics[width=\columnwidth]{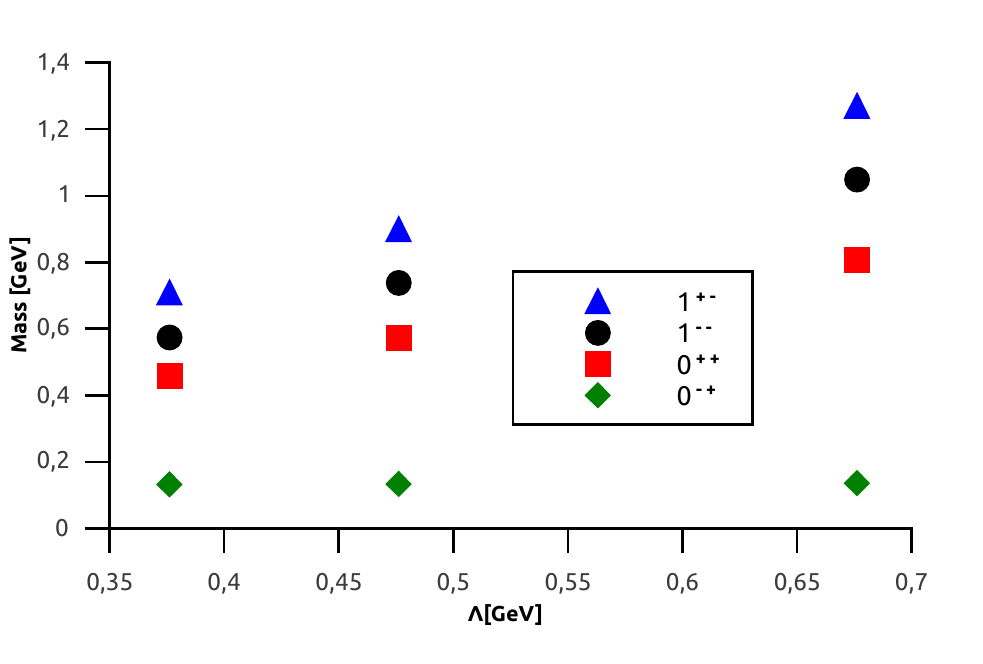}
\caption{This figure shows the dependence of the ground state meson masses on the 
details of the quark-gluon interaction with Munczek vertex. 
In the upper panel $\Lambda=0.376$ is held fixed, in the lower panel $\eta=1.315$. }
\label{fig:Mun_results2}       
\end{figure}

We investigated the Munczek vertex for a similar behavior. In order to assess the model 
dependence of our results we varied the parameters $\eta$ and $\Lambda$ in the ansatz
for the effective interaction, see Eq.~(\ref{eqn:MarisTandyInteraction}). For each value 
of the parameters we adjusted the bare quark mass to obtain roughly the physical
pion mass of $m_\pi=0.137\,$GeV. We then calculated the pion decay constant on the pion 
mass shell, see \cite{Maris:1997hd} for numerical details, and the masses for the pion, 
the scalar, the vector and the axial-vector ground states. Our results are shown in 
Fig.~\ref{fig:Mun_results2} and in Tab.~\ref{tab:Mun_result_table}. 

In general, the pion mass serves to fix the input quark mass, whereas the pion decay constant
is sensitive to the scale of the interaction. In pure rainbow ladder calculations it has been 
observed that once the scale is fixed via $\Lambda$, there is a whole range of values
for $\eta$ which leave the pion decay constant untouched. It has also been established, that 
the masses of the scalar and vector meson bound states are almost insensitive to these
variations \cite{Krassnigg:2009zh}. This is no longer true, when the vertex is non-trivial
as can be seen from Fig.~\ref{fig:Mun_results2}. Varying the parameters of the interaction 
one clearly finds a great impact onto the meson mass spectrum. This comes with an increase
of the splitting between pseudoscalar and scalar channels as well as vector and axialvector 
channels. Thus in principle, by variation of the model parameters one could drive the masses
of the scalar and axialvector states in a region around and above 1 GeV, where they could be 
identified with physical states such as the $f_0(1370)$ and the $a_1(1260)$.

\begin{table}[t]
 \centering
 \begin{center}
\begin{tabular}{|l|l|l||l||l|l|l|}\hline
$\Lambda$ [GeV]\hspace*{-1mm} & $\eta$   & \hspace*{-1mm}f$_\pi$ [GeV]\hspace*{-1mm}&\hspace*{-1mm}&
$\Lambda$ [GeV]\hspace*{-1mm} & $\eta$   & \hspace*{-1mm}f$_\pi$ [GeV]\hspace*{-1mm}                         \\\hline
 0.376     & 1.315    & 0.094         && 0.376     & 1.315    & 0.094         \\\hline
 0.376     & 1.415    & 0.103         && 0.476     & 1.315    & 0.118         \\\hline
 0.376     & 1.515    & 0.109         && 0.676     & 1.315    & 0.165         \\\hline
 \end{tabular}
 \end{center}
 \caption{Results for the pseudoscalar decay constants for different parameters in the  
 quark-gluon interaction with Munczek vertex.}
  \label{tab:Mun_result_table}
\end{table}

\begin{table}[htp]
 \centering
 \begin{center}
\begin{tabular}{|l|l|l|l|l|l|l|l|}\hline
   \hspace*{-1mm}& $\Lambda$ [GeV]\hspace*{-1mm}&  $\eta\hspace*{-1mm}$  & $f_\pi$ &  $m_\pi$ & $m_\sigma$  & $m_\rho$ & $m_{a1}$     \\\hline
\hspace*{-1mm}RL & 0.7     & 1.8      & 0.093   & 0.137    & 0.65    & 0.73     &  0.83	  \\\hline
\hspace*{-1mm}MV & 0.376   & 1.315    & 0.094   & 0.134    & 0.46    & 0.58     &  0.71	  \\\hline

 \end{tabular}
 \end{center}
 \caption{Meson masses and decay constants (in units of GeV) for Rainbow ladder (RL) 
 compared with our results using the Munczek vertex (MV).
 }
  \label{tab:Mun_result_table2}
\end{table}

However, with the construction at hand this would be stretching the model 
much too far: as can be seen from Tab.~\ref{tab:Mun_result_table} also the
pion decay constant increases with increased spin-orbit splitting, clearly
indicating that one is no longer working with acceptable model parameters.
Indeed, when we compare the rainbow-ladder result (RL) with the improved 
approximation scheme using the Muczek vertex (MV) in 
Tab.~\ref{tab:Mun_result_table2} with model parameters adjusted such that 
the pion decay constant comes out right we even observe a decrease of the
spin-orbit splitting. 
Similar results can be obtained with the improved
Ball-Chiu vertex (ABC) of section \ref{sec:BCVertex}. We thus find, that
a Ward-Identity improved vertex alone is not enough to reproduce the 
size of the spin-orbit splitting that is suggested from 
experiment. Note that we do not put much emphasis on the fact, that the 
mass of the quark-antiquark bound state in the scalar channel using the
MV-vertex is even in the right ballpark for the $f_0(500)$. As noted in
Ref.~\cite{Chang:2011ei} there are indeed transverse parts of the vertex 
that do increase the spin-orbit splitting by a substantial amount thus 
making the identification with the $f_0(1370)$ more likely. This also ties 
in with findings of Ref.~\cite{Heupel:2012ua}.

\section{Conclusions}
\label{sec:conclusions}

Following the time-honored concept of taking functional derivatives to obtain an 
interaction kernel, we extended this technique to vertex models which 
explicitly depend on the quark propagator and it's dressing functions. 
This enabled us to derive closed expressions for the interaction kernel
beyond the rainbow-ladder approximation. Our technique is very general,
and in principle applicable to any vertex that is given in terms of quark
dressing functions. As an improvement over previous approaches 
\cite{Chang:2009zb,Chang:2010hb,Chang:2011ei} our technique allows to
determine not only the masses of the bound states but also their 
Bethe-Salpeter wave functions. Certainly, these are indispensable when 
it comes to the calculations of form factors, structure functions, or 
decay widths of the states in question. 

As examples, we applied this technique to two type of vertices, the 
Ball-Chiu vertex and the Munczek vertex that both respect the constraints 
due to the vector Ward-Takahashi identity. For the Ball-Chiu vertex we 
find that we have to amend the vertex by additional parts along unphysical
directions in Dirac space. These do not contribute to the Dyson-Schwinger
equation for the quark propagator, but generate important additional terms 
into the interaction kernel of Bethe-Salpeter equations necessary to respect
the axial Ward-Takahashi identity. The resulting pion is then a Goldstone 
boson in the chiral limit. For the Munczek vertex, such additional
contributions are not necessary. 

Using the Munczek vertex we performed a calculation of the masses of 
pseudoscalar, scalar, vector and axialvector mesons and confirm the 
findings of Ref.\cite{Chang:2009zb}: the additional gauge related 
structure in the vertex is dominated by dynamical effects of chiral 
symmetry breaking and capable to generate substantial spin-orbit 
forces. However, these structures alone are not sufficient to generate 
a physical spectrum of light mesons while keeping the pion properties
intact. Additional transverse pieces in the vertex are necessary to
improve this situation.  

\vspace*{5mm}
\noindent{\bf Acknowledgments}\\
We are grateful to Richard Williams and Gernot Eichmann for discussions
and comments on the manuscript.
This work was supported by BMBF under contract \linebreak 06GI7121, by 
the Helmholtz International Center for FAIR within the LOEWE 
program of the State of Hesse and by the Helmholtzzentrum GSI.

\appendix
\section{Constructing beyond ladder kernels}
\label{app:Kernels}
Here we detail the construction of type $II$ kernels
in order to provide a self-contained definition
that should help the reader who is interested 
in the numerical implementation. 
We consider in particular the $ABC$
vertex construction from Eq. (\ref{eqn:ABCVertex}).
The quark dressing functions are taken to be the ones
from Eq. (\ref{eqn:unphysQuark}).

The kernels are of the form
\begin{align}
  \left.\frac{\delta \Sigma^{ab}(k)}{\delta S^{cd}(q)}\right|_{II}=
  \int_l [\gamma_\mu S(l)]^{aa^\prime} \frac{\delta \Gamma_\mu^{a^\prime b}(l,k)}{\delta S^{cd}(q)}D_{\mu\nu}(l-k),
  \label{eqn:BeyondLadderKernelAGAIN}
\end{align}
where the vertex from Eq. (\ref{eqn:ABCVertex}) using
the generalisation from Eq. (\ref{eqn:2BCunphysical}) is written as
\begin{align}
  &\Gamma_\mu^{ABC}(l,k) = 
  \gamma_\mu\frac{A_\mu(l)+A_\mu(k)}{2}\notag\\
  &+(l+k)_\mu\sum_{\alpha=1}^4 (l+k)_\alpha\gamma_\alpha\frac{1}{2}\frac{A_\alpha(l)-A_\alpha(k)}{l^2-k^2}
    \label{eqn:ABCVertexUnphys}\\
  &+i\,(l+k)_\mu \frac{B(l)-B(k)}{l^2-k^2}
  +i\,\gamma_5\,(l+k)_\mu \frac{C(l)-C(k)}{l^2-k^2}\notag,
 \end{align}
which is the analog of the Ball-Chiu construction for
the quark shown in Eq. (\ref{eqn:unphysQuark}).
The cutting is now explicitly done as
\begin{align}
  \frac{\delta}{\delta S^{cd}(q)}= 
 \sum_{j=1}^4\frac{\gamma^{dc}_j}{i\,4\,q_j}  \frac{\delta }{\delta \sigma_j(q)}  
 +\frac{\OperatorOne^{dc}}{4} \frac{\delta}{\delta \sigma_S(q)}
 +\frac{\gamma_5^{dc}}{4}\frac{\delta}{\delta \sigma_5(q)}
  \label{eqn:CuttingOperator}
\end{align}
The functional derivatives that occur are of the form
\begin{align}
  \frac{\delta A_i(p)}{\delta \sigma_j(q)} = \frac{\partial A_i}{\partial \sigma_j}(p)\, \delta^{(4)}(p-q),
  \label{eqn:DefineParitalDerivative}
\end{align}
and similar for the $B$ and $C$ functions. We need to
specify $A$, $B$ and $C$ in terms of the $\sigma$-dressings.
The quark and its' inverse defined as in
Eq. (\ref{eqn:unphysQuark}) are related by
\begin{xalignat}{2}
  A_i &=\frac{\sigma_i}{\sum_i^4 p_i^2\sigma_i^2 + \sigma_S^2-\sigma_5}&\rightarrow\quad
  &\frac{\sigma_V}{p^2\sigma_V^2+\sigma_S^2}\notag\\
  B   &=\frac{\sigma_S}{\sum_i^4 p_i^2\sigma_i^2 + \sigma_S^2-\sigma_5}&\rightarrow\quad
  &\frac{\sigma_S}{p^2\sigma_V^2+\sigma_S^2}\label{eqn:ABCexpressions}\\
  C &=-\frac{\sigma_5}{\sum_i^4 p_i^2\sigma_i^2 + \sigma_S^2-\sigma_5}&\rightarrow\quad
   & 0\notag,
\end{xalignat}
where the expressions after '$\rightarrow$' are the ones after $\sigma_{1\dots4}\rightarrow\sigma_V$
and $\sigma_5\rightarrow 0$, i.e. the physical ones that are used in all numerical calculations.
The 'unphysical' expressions in Eq. (\ref{eqn:ABCexpressions}) are only needed for the cutting
procedure during the derivation of the type $II$ kernels. The coefficient matrix from
Eq. (\ref{eqn:DefineParitalDerivative}) evaluates to
\begin{align}
  \frac{\partial (A_i|B|C)}{\partial \sigma_j}&=\frac{1}{\mathcal{N}^2}
 \begin{bmatrix}
   D_1 & \Sigma^V_2 &\Sigma^V_3 & \Sigma^V_4 & \Sigma^{VS}& 0 \\
   \Sigma^V_1 & D_2 &\Sigma^V_3 &\Sigma^V_4 & \Sigma^{VS} & 0\\
   \Sigma^V_1 &\Sigma^V_2 & D_3 &\Sigma^{V}_4 & \Sigma^{VS}& 0\\
   \Sigma^V_1  & \Sigma^V_2  & \Sigma^V_3  & D_4 & \Sigma^{VS}& 0\\
   \Sigma^{VS}_1 & \Sigma^{VS}_2 & \Sigma^{VS}_3 & \Sigma^V_4 & D_S & 0\\
   0 & 0 & 0 & 0 & 0 & -\mathcal{N}
 \end{bmatrix},
 \label{eqn:CoefficientMatrix}
\end{align}
with
\begin{xalignat}{2}
  \mathcal{N}& = p^2\sigma_V^2+\sigma_S^2 &   D_i& = \sigma_V^2\bigg( \big(\sum_{j\neq i}p_j^2\big)-p_i^2\bigg)+\sigma_S^2\notag\\
  D_S &= p^2\sigma_V^2 - \sigma_S^2 & \Sigma^V_i &= -2\sigma_V^2 p_i^2\label{eqn:CoeffMatSupplement}\\
  \Sigma^{VS}_i & = -2 \sigma_V\sigma_S p_i^2 &  \Sigma^{VS} & = -2 \sigma_V \sigma_S\notag.
\end{xalignat}
Note that momentum $p$ in the equations above will be evaluated as $l$ or $k$ 
in equation (\ref{eqn:ABCVertexUnphys}).
The type $II$ kernel for the vertex model from Eq. (\ref{eqn:ABCVertex}) is now
almost fully specified. In addition we adjust the momentum dependence in order 
to take into account the flow of the total momentum of the bound state through 
the kernel. This procedure is explained in appendix \ref{app:AXWTI3BC} for the 
case of the $\delta C/\delta \sigma_5$ part.

\section{Massless pion and 3BC vertex}
\label{app:AXWTI3BC}

In this appendix we show how the Ball-Chiu vertex
(Eq. (\ref{eqn:3rdBCVertex}))
can yield a massless pion in the chiral limit via
the extended structure of the vertex from Eq. (\ref{eqn:ABCVertex}).
The only type $II$ term in the kernel originating from cutting
the $ABC$ vertex of Eq. (\ref{eqn:ABCVertex}) and contributing to 
the AXWTI (\ref{eqn:AXWTI2}) and the pion BSE (\ref{eqn:PionBSE})
evaluates to (see appendix \ref{app:Kernels})
\begin{eqnarray}
  \frac{\delta \Sigma^{cd}(k)}{\delta S^{ab}(q)}&=&-\gamma_5^{ba}/4
  \label{eqn:TypeIIkernel}\\
  &&\hspace*{-5mm}\times\Bigg[[\gamma_\mu S(q)\gamma_5]^{cd}\frac{i (q+k)_\nu}{q^2-k^2}\frac{D_{\mu\nu}(q-k)}{q^2\sigma_V^2(q^2)+\sigma_S^2(q^2)}   \nonumber\\
  &&\hspace*{-5mm}-\int_l[\gamma_\mu S(l)\gamma_5]^{cd}\frac{i (l+k)_\nu}{l^2-k^2}\frac{D_{\mu\nu}(l-k)\,\delta^{(4)}(q-k)}{k^2\sigma_V^2(k^2)+\sigma_S^2(k^2)} \Bigg].\nonumber
\end{eqnarray}
We mentioned already in section \ref{sec:General} that the kinematics 
of the kernels generated is not automatically given by the cutting procedure. 
The self-energy $\Sigma(k)$ expects the same incoming and outgoing momenta. 
The kernel that is generated from its derivative should have different 
momenta $k_+$ and $k_-$ to match the kinematics needed in the bound state 
equation (\ref{eqn:PionBSE}). If the cutting were carried out in coordinate 
space, this ambiguity would not arise. In order to arrive at a fully specified
kernel one should use Eq.~(\ref{eqn:KasFuntDerivative}) without assuming 
translational invariance but only relaxing all Green functions to physical 
ones in the end. We were, however, so far unable to write down the 
Ball-Chiu construction (\ref{eqn:3rdBCVertex}) for a quark with different 
in- and out-going momenta, or, probably preferable, in coordinate space.
In our numerical calculation we thus work with a momentum-shifted version 
of Eq.~(\ref{eqn:TypeIIkernel}) which reads
\begin{eqnarray}
  K^{ABC}_{II}(P,q,k)_{ab}^{cd}&=&-\gamma_5^{ba}/4 \label{eqn:TypeIIkernelShifted}\\
  &&\hspace*{-20mm}\times\Bigg[[\gamma_\mu S(q)\gamma_5]^{cd}\frac{i (q+k_+)_\nu}{q^2-k_+^2}\frac{D_{\mu\nu}(q-k)}{q^2\sigma_V^2(q^2)+\sigma_S^2(q^2)} \nonumber  \\
 &&\hspace*{-20mm} - \int_l[\gamma_\mu S(l)\gamma_5]^{cd}\frac{i (l+k_+)_\nu}{l^2-k_+^2}\frac{D_{\mu\nu}(l-k)\,\delta^{(4)}(q-k)}{k_+^2\sigma_V^2(k_+^2)+\sigma_S^2(k_+^2)} \Bigg]\,.\nonumber
\end{eqnarray}
Our reasoning for this expression is twofold. First of all it does respect 
the fact that the total momentum $P$ that should be part of the kernel, as 
explained above. Second, the singular terms of the form $1/(q^2-k^2)$ are 
potentially dangerous in the integration. This is regularized due to the 
replacement $k\rightarrow k_+=k+P/2$, where $P$ is imaginary ($P^2=-m_\pi^2$).
It turns our that with this momentum routing a cancellation between the two 
types of structures present in Eq.~(\ref{eqn:TypeIIkernelShifted}) occurs. 
This cancellation mechanism resembles the vertex structure from 
Eq.~(\ref{eqn:ABCVertex}), where the same type of denominator occurs. However,
since the quotient approaches a form that is reminiscent of a derivative 
$dC(k^2)/dk^2$ the zero-momentum limit is well defined.

We will now show that the AXWTI, Eq.~(\ref{eqn:AXWTI2}), is fulfilled in 
the limit of $P\rightarrow 0$. For the case of the type I contribution to
the kernel, only the third term of the Ball-Chiu vertex, Eq.~(\ref{eqn:3rdBCVertex}),
is a problem (cf. section \ref{sec:2BCVertex}). This is because the $\gamma_5$'s 
on the right hand side of equation (\ref{eqn:AXWTI2}) have to anti-commute 
with the vertices to give an additional minus sign to match the left side 
of the equation (cf. Fig.~\ref{fig:AXWTI}). This works out for the first two 
components of the BC vertex: $\{\Gamma_\mu^{2BC},\gamma_5\}=0$. The third 
component generates a term with the wrong sign since $[\Gamma_\mu^{3rdBC},\gamma_5]=0$.
It turns out, however, that the type II contributions to the kernel,
Eq.~(\ref{eqn:BRLKernelC5}), remedy the problem: they equal to twice the same
contribution but with opposite sign and therefore effectively switch the sign.

In order to be explicit we will start to check the AXWTI, Eq.~(\ref{eqn:AXWTI2}) 
for the case of a bare vertex $\Gamma_\mu(q,p)=\gamma_\mu$ in 
Eq.~(\ref{eqn:LadderKernel}), i.e. the rainbow-ladder case. The essential 
manipulation in Eq.~(\ref{eqn:AXWTI2}) is
\begin{align}
  -\int_q [&S(q_+)\gamma_5]^{cd}K^{ab}_{cd}(P,q,k)\label{eqn:CheckAXWTIforRL}\\
  & = -\int_q \gamma_\mu S(q_+)\gamma_5 \gamma_\nu D_{\mu\nu}(k-q)\notag\\
 & = \int_q \gamma_\mu S(q) \gamma_\nu D_{\mu\nu}(k_+-q)\gamma_5=\Sigma(k_{+})\gamma_5,\notag
\end{align}
which then matches a corresponding term on the left side of equation (\ref{eqn:AXWTI2}).
For the case of a generic vertex that fulfills $\{\Gamma_\mu(q,k),\gamma_5\}=0$ we find
\begin{align}
  -\int_q [&S(q_+)\gamma_5]^{cd}K^{ab}_{cd}(P,q,k)\label{eqn:CheckAXWTIfor2BC}\\
  & = -\int_q \gamma_\mu S(q_+)\gamma_5 \Gamma_\nu(q_-,k_-) D_{\mu\nu}(k-q)\notag\\
 & = \int_q \gamma_\mu S(q) \Gamma_\nu(q_--P/2,k_-) D_{\mu\nu}(k_+-q)\gamma_5\notag\\%=\Sigma(k_{+})\gamma_5,\notag\\
 &\stackrel{P\rightarrow 0}{\rightarrow}
   \int_q \gamma_\mu S(q) \Gamma_\nu(q,k) D_{\mu\nu}(k-q)\gamma_5=\Sigma(k)\gamma_5,\notag
\end{align}
such that the AXWTI is fulfilled at $P=0$. For a vertex component, such as 
the third term of the BC part in Eq. (\ref{eqn:3rdBCVertex}) that obeys 
$[\Gamma_\mu,\gamma_5]=0$ the contribution has the wrong sign, such that 
even at $P=0$ the AXWTI is not fulfilled.

We will see that this problem can be cured by including a contribution of 
type $II$. Using the definition $f_\nu(q,k)= (q+k)_\nu/(q^2-k^2)$, the 
self-energy for the third BC component on the left hand side of the AXWTI reads
\begin{align}
  \int_q\gamma_\mu S(q)f_\nu(q,k_+)\big(B(q)-B(k_+)\big)D_{\mu\nu}(k_+-q)\gamma_5.
  \label{eqn:3BCSelfEnergy}
\end{align}
The corresponding diagram on the right side of the AXWTI has the opposite 
sign as already stated above. Therefore we consider now the contribution of 
the type $II$ kernel from Eq.~(\ref{eqn:TypeIIkernel}). The corresponding $C$ 
part of the vertex (\ref{eqn:ABCVertex}) is zero and does not contribute
to the self-energies on the left side of the AXWTI. For simplicity we will 
use the function $f_\nu$ again and also the function $\mathcal{N}$ from 
Eq.~(\ref{eqn:CoeffMatSupplement}).
%
%
%\pagebreak
%\begin{widetext}
\begin{align}
  &-\int_q [S(q_+)\gamma_5]^{cd}K^{ab}_{cd}(P,q,k)
  =\frac{1}{4}\int_q \mathrm{Tr}\left[ S(q_+)\gamma_5\gamma_5 \right]\times\notag\\
  &\hspace*{15mm}\Bigg[\gamma_\mu S(q)\gamma_5\frac{f_\nu(q,k_+)}{\mathcal{N}(q)}D_{\mu\nu}(q-k)\notag\\
 & \hspace*{15mm}-\int_l \gamma_\mu S(l)\gamma_5\frac{f_\nu(l,k_+)}{\mathcal{N}(k_+)}D_{\mu\nu}(l-k)\delta(q-k)\Bigg]\notag\\
 &=\int_q\gamma_\mu S(q)f_\nu(q,k_+)\frac{\sigma_S(q_+)}{\mathcal{N}(q)}D_{\mu\nu}(q-k)\gamma_5\notag\\
 &\hspace*{15mm}- \int_l\gamma_\mu S(l) f_\mu(l,k_+)\frac{\sigma_S(k_+)}{\mathcal{N}(k_+)}D_{\mu\nu}(l-k)\gamma_5\notag\\
 &=\int_q\gamma_\mu S(q)f_\mu(q,k_+)\left[\frac{\sigma_S(q_+)}{\mathcal{N}(q)}
 -  \frac{\sigma_S(k_+)}{\mathcal{N}(k_+)}  \right]D_{\mu\nu}(k-q)\gamma_5\notag\\
 &\stackrel{P\rightarrow 0}{\rightarrow}\int_q\gamma_\mu S(q)f_\mu(q,k)
 (B(q)-B(k))D_{\mu\nu}(k-q)\gamma_5.
 \label{eqn:RHSAXWTIforC5}
\end{align}
%\end{widetext}
Here $\mathrm{Tr}[S]=4 \sigma_S$ was used as well as the definition of 
the $B$ function in Eq. (\ref{eqn:ABCexpressions}).
We see that the last line corresponds to Eq. (\ref{eqn:3BCSelfEnergy})
in the $P\rightarrow 0$ limit. In fact the second contribution on the right side
of the AXWTI (\ref{eqn:AXWTI2}) differs by $S(k_+)\gamma_5 \rightarrow \gamma_5S(k_-)$
such that in the $P\rightarrow 0$ limit it yields the same contribution.
Thus we have the contribution of Eq. (\ref{eqn:RHSAXWTIforC5})
twice. Due to the global minus sign that comes from the definition
of $C$ in Eq. (\ref{eqn:ABCexpressions}) we subtract the 
3BC term from Eq. (\ref{eqn:3BCSelfEnergy}) twice such
that the AXWTI is fulfilled in the $P\rightarrow 0$ limit.

\section{Gluon model}
\label{app:gluon}
In this work we use a model for the effective gluon propagator ${D}_{\mu\nu}$ that 
was given in Ref.~\cite{Maris:1999nt}.
In general the gluon is given in Landau gauge as
\begin{align}
  \tilde{D}_{\mu\nu}(k) 
  =\left( \delta_{\mu\nu}-\frac{k_\mu k_\nu}{k^2} \right)\frac{Z(k^2)}{k^2},
  \label{eqn:GLuonPropagator}
\end{align}
where the non perturbative content is hidden in the dressing function $Z(k^2)$.
In the Dyson-Schwinger equation for the quark propagator this dressing function
appears together with the fully dressed non-Abelian quark-gluon vertex. Since all 
explicit vertices used in this work are constructed along the Abelian Ward-Takahashi 
identity, the following model for the effective gluon represents a product of the 
gluon propagator with the remaining non-Abelian dressing effects $G\tilde{H}$
in the vertex, cf. the discussion around Eq.~(\ref{eqn:3rdBCVertex}). 
The model is given by 
\begin{flalign}\label{eqn:MarisTandyInteraction}
\alpha_{\textrm{eff}}(k^2) &= \frac{g^2}{4\pi} Z_{1F} Z(k^2) G(k^2) \tilde{H}(k^2)
\notag\\
&=
 \pi\eta^7\left(\frac{k^2}{\Lambda^2}\right)^2
e^{-\eta^2\frac{k^2}{\Lambda^2}}\nonumber\\ &\hspace*{15mm}+{}\frac{2\pi\gamma_m
\big(1-e^{-k^2/\Lambda_{t}^2}\big)}{\textnormal{ln}[e^2-1+(1+k^2/\Lambda_{QCD}
^2)^2]}\,, 
\end{flalign}
where for the anomalous dimension of the quark we use $\gamma_m=12/(11N_c-2N_f)=12/25$,
corresponding to $N_f=4$ flavors and $N_c=3$ colors, we fix the QCD scale to
$\Lambda_{QCD}=0.234$ GeV and the scale $\Lambda_t=1$~GeV is introduced for 
technical reasons and has no impact on the results. The interaction strength 
is characterized by an energy scale $\Lambda$ and the dimensionless parameter
$\eta$ controls the width of the interaction.
The precise form of this model does not matter in this work. Ultimately
we aim to replace this with a self-consistently calculated gluon propagator, 
see e.g. Ref.~\cite{Fischer:2003rp}, and an appropriate expression for the
non-Abelian parts of the vertex. 

\section{Munczek model}
\label{app:MunModel}
\begin{figure}[b]
      \centering
\resizebox{0.35\textwidth}{!}{%
  \includegraphics{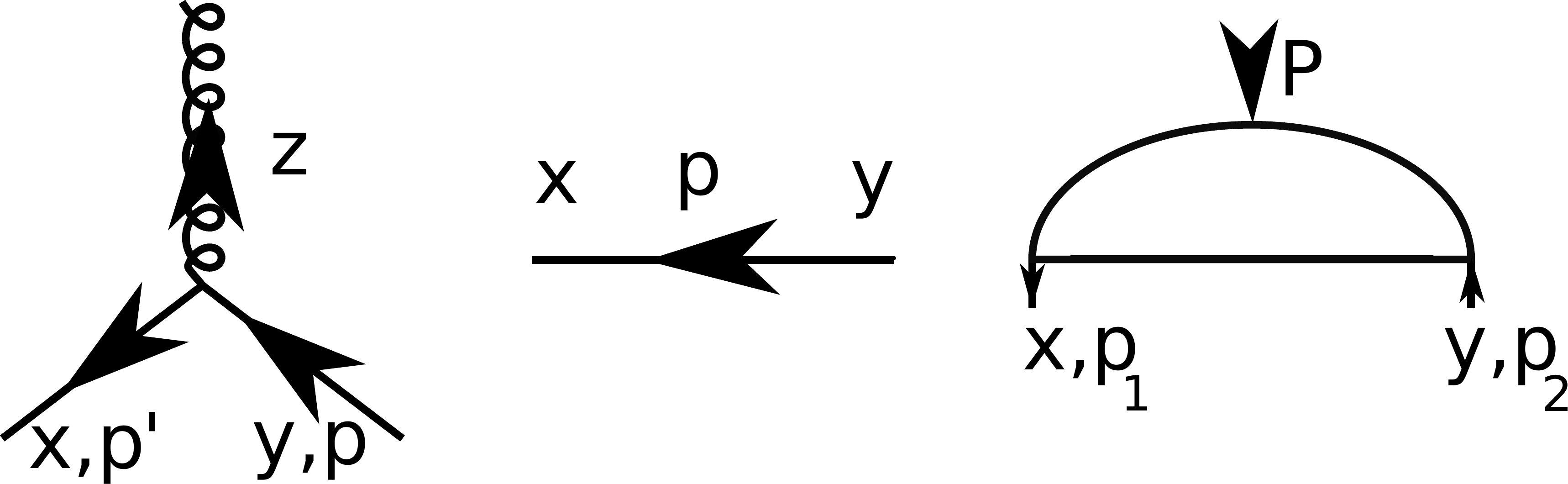}
}    
       \caption{From left to right: Vertex, Propagator, Amplitude.
                $x,y,z$ letters denote spacetime positions, $p$ letters denote the corresponding momentum.}
 \end{figure}
To cut the quark selfenergy of the Munczek model, a formulation in position space is necessary. The building blocks read as following:
\begin{align}
  \Gamma^{\mu}\left(z;x,y\right)&=\int\limits_{p^\prime,p}e^{-ip^\prime(x-z) - ip(z-y)}\,\Gamma^{\mu}\left(p^{\prime},p\right)\nonumber\\
  S\left(x,y\right) /D^{\mu\nu}\left(x,y\right)&= \int\limits_p e^{-ip(x-y)}\, S/D^{\mu\nu}\left(p\right)\nonumber\\
  A\left(x,y;P\right)&=\int\limits_{p_1,p_2}e^{-ip_1x+ip_2y}\,A(p_1,p_2;P).\label{eq:Amplitude_cs}
\end{align}
The first line represents a vertex in position space, the second a  quark or gluon propagator and the third one the Bethe-Salpeter amplitude or wavefunction with total momentum $P$.
We denote two further relations that play a role in the derivations:
\begin{gather}
 \int\limits_0\limits^1d\alpha \,e^{iq\alpha (x-y)} e^{iq(z-x)}=
 \frac{e^{iq(z-y)} - e^{iq(z-x)}}{iq\cdot(x-y)}\label{eq:alpha_trick},
\end{gather}
and 
\begin{gather}
 \frac{\delta}{\delta S(l,l^\prime)}S^{-1}(x^{\prime},y)= - S^{-1}(l^{\prime},y)S^{-1}(x^\prime,l).\label{eq:S1_Cut}
\end{gather}
The first one is  already used to represent the Munczek vertex model in \cite{Munczek:1994zz}, the second one denotes the functional derivative of an inverse propagator.
With all tools at hand the Munzcek vertex in momentum space is readily derived from 
Eq.~(\ref{eqn:MunczekVertex}):
\begin{align}
 \Gamma^{\mu}(p^\prime,p)&=\frac{\partial}{\partial p_\mu}\int\limits^1\limits_0  S^{-1}(p+\alpha (p^\prime - p))\, d\alpha.
\end{align}
Taking the functional derivative of the quark selfenergy yields
\begin{align}
 \frac{\delta\Sigma(x_1,x_2)}{\delta S(l,l^\prime)}	&= 	\int\limits_{y,z}\gamma^\mu S(x_1,y)\frac{\delta\Gamma^{\mu}(z;y,x_2)}{\delta S(l,l^\prime)}D^{\mu\nu}(z,x_1)
 \nonumber\\
 &=\int\limits_{y,z}\gamma^\mu S(x_1,y) D^{\mu\nu}(z,x_1)\int\limits_q 
 e^{iq(z-y)-iq(z-x_2)} \nonumber\\
 &\hspace*{5mm}\times \frac{\left(x_2-y\right)^\mu}{iq\cdot(x_2-y)}\frac{\delta}{\delta S(l,l^\prime)}S^{-1}(y,x_2).\nonumber
\end{align}
This expression is now traced with the Bethe-Salpeter wave function from Eq.\eqref{eq:Amplitude_cs} (as demanded by the Bethe-Salpeter Equation in coordinate space)  and Eq.\eqref{eq:S1_Cut} is inserted for the derivative of the inverse quark propagator. Additionally the $\alpha$-trick
from Eq.\eqref{eq:alpha_trick} is applied resulting in the following expression
\begin{align}
  \frac{\delta\Sigma(x_1,x_2)}{\delta S(l,l^\prime)}&=-\int\limits_{\substack{y,z\\l,l^\prime,q}}\int\limits_0\limits^1d\alpha\, \gamma^\mu S(x_1,y)D^{\mu\nu}(z,x_1)e^{iq\alpha(x_2-y)}\nonumber\\
 &\hspace*{-10mm}\times\,e^{iq(z-x_2)}\left(x_2-y\right)^\mu S^{-1}(l^\prime,x_2)\tilde{\Gamma}(l^\prime,l;P)S^{-1}(y,l),
\end{align}
where $\tilde{\Gamma}$ is the wave function (see Eq. (\ref{eqn:WaveFunction})).
Inserting  the expressions for the vertex, propagators and amplitudes from Eq.\eqref{eq:Amplitude_cs} and replacing $\left(x_2-y\right)^\mu$ by appropriate derivatives of momenta in the exponentials of the Fourier modes,
one finally arrives at the expression for the type $II$ momentum space contribution:
\begin{align}
 \left[\Gamma\times K_{II}\right](P,p)|_{ab}&=-\int\limits_{\substack{q\\}}\int\limits_0\limits^1d\alpha\,  
\gamma_{ac}^\mu S_{cd}(q -\frac{1}{2} P) \nonumber\\ 
&\hspace*{-10mm}\times D^{\mu\nu}(q-p)\left[ \frac{\partial}{\partial q^\nu}\left( \Gamma_{db}(q+\alpha (q-p);P)\right)\right]. 
\end{align}
Color factors and renormalization constants are suppressed.
In this case $\Gamma$ denotes the wave function.
and instead of the two momenta $p_1$ and $p_2$, we use the relative momentum $p=(p_1-p_2)/2$ to describe the wave function. 
We included the Dirac indices to clarify the structure.\\
There is an asymmetry in this type $II$ kernel as one can see in the quark momentum. This can
lead to an imaginary part of the BSE eigenvalues at least in the form of numerical noise.
The source for this is the asymmetry
of the quark self-energy that contains only one dressed vertex. If one would start with a symmetrised self-energy
\begin{gather}
  \Sigma(p)=\frac{1}{2}\int\, \gamma^\mu S(q)\Gamma^\nu(q,p) D^{\mu\nu}(p-q) \\
   + \frac{1}{2}\int\, \Gamma^\mu(p,q) S(q)\gamma^\mu D^{\mu\nu}(p-q)
 \label{eq:Sym_Self}
\end{gather}
 this problem disappears and there is second type $II$ contribution containing a quark with momentum $q+\frac{1}{2}P$:
\begin{gather}
 \left[\Gamma\times K_{II}\right](P,p)|_{ab}=-\int\limits_{\substack{q\\}}\int\limits_0\limits^1d\alpha\,  
   \left[ \frac{\partial}{\partial q^\nu}\left( \Gamma_{ac}(q+\alpha (q-p);P)\right)\right]\nonumber\\ 
\times S_{cd}(q +\frac{1}{2} P)\gamma_{db}^\mu  D^{\mu\nu}(q-p). 
\end{gather} 
Both contributions will come with a factor $\frac{1}{2}$.


\begin{thebibliography}{}

%\cite{McKay:1989rk}
\bibitem{McKay:1989rk} 
  D.~W.~McKay and H.~J.~Munczek,
  %``Composite Operator Effective Action Considerations on Bound States and Corresponding S Matrix Elements,''
  Phys.\ Rev.\ D {\bf 40}, 4151 (1989).
  %%CITATION = PHRVA,D40,4151;%%

\bibitem{Munczek:1994zz} 
  H.~J.~Munczek,
  %``Dynamical chiral symmetry breaking, Goldstone's theorem and the consistency of the Schwinger-Dyson and Bethe-Salpeter Equations,''
  Phys.\ Rev.\ D {\bf 52}, 4736 (1995)
  [hep-th/9411239].
  %%CITATION = HEP-TH/9411239;%%

%\cite{Maris:1997hd}
\bibitem{Maris:1997hd} 
  P.~Maris, C.~D.~Roberts and P.~C.~Tandy,
  %``Pion mass and decay constant,''
  Phys.\ Lett.\ B {\bf 420}, 267 (1998)
  [nucl-th/9707003].
  %%CITATION = NUCL-TH/9707003;%%

%\cite{Maris:2003vk}
\bibitem{Maris:2003vk}
  P.~Maris and C.~D.~Roberts,
  %``Dyson-Schwinger equations: A Tool for hadron physics,''
  Int.\ J.\ Mod.\ Phys.\ E {\bf 12} (2003) 297
  [nucl-th/0301049].
  %%CITATION = NUCL-TH/0301049;%%

%\cite{Maris:2005tt}
\bibitem{Maris:2005tt}
  P.~Maris and P.~C.~Tandy,
  %``QCD modeling of hadron physics,''
  Nucl.\ Phys.\ Proc.\ Suppl.\  {\bf 161} (2006) 136
  [nucl-th/0511017].
  %%CITATION = NUCL-TH/0511017;%%

%\cite{Goecke:2011pe}
\bibitem{Goecke:2011pe}
  T.~Goecke, C.~S.~Fischer and R.~Williams,
  %``Leading-order calculation of hadronic contributions to the muon $g-2$ using the Dyson-Schwinger approach,''
  Phys.\ Lett.\ B {\bf 704} (2011) 211
  [arXiv:1107.2588 [hep-ph]];
  %%CITATION = ARXIV:1107.2588;%%
%\cite{Goecke:2012qm}
%\bibitem{Goecke:2012qm}
%  T.~Goecke, C.~S.~Fischer and R.~Williams,
  %``The role of momentum dependent dressing functions and vector meson dominance in hadronic light-by-light contributions to the muon $g-2$,''
  Phys.\ Rev.\ D {\bf 87} (2013) 3,  034013
  [arXiv:1210.1759 [hep-ph]].
  %%CITATION = ARXIV:1210.1759;%%

%\cite{Eichmann:2009qa}
\bibitem{Eichmann:2009qa}
  G.~Eichmann, R.~Alkofer, A.~Krassnigg and D.~Nicmorus,
  %``Nucleon mass from a covariant three-quark Faddeev equation,''
  Phys.\ Rev.\ Lett.\  {\bf 104} (2010) 201601
  [arXiv:0912.2246 [hep-ph]].
  %%CITATION = ARXIV:0912.2246;%%

%\cite{Eichmann:2008ae}
\bibitem{Eichmann:2008ae}
  G.~Eichmann, R.~Alkofer, I.~C.~Cloet, A.~Krassnigg and C.~D.~Roberts,
  %``Perspective on rainbow-ladder truncation,''
  Phys.\ Rev.\ C {\bf 77} (2008) 042202
  [arXiv:0802.1948 [nucl-th]].
  %%CITATION = ARXIV:0802.1948;%%

%\cite{Qin:2011xq}
\bibitem{Qin:2011xq}
  S.~-x.~Qin, L.~Chang, Y.~-x.~Liu, C.~D.~Roberts and D.~J.~Wilson,
  %``Investigation of rainbow-ladder truncation for excited and exotic mesons,''
  Phys.\ Rev.\ C {\bf 85} (2012) 035202
  [arXiv:1109.3459 [nucl-th]].
  %%CITATION = ARXIV:1109.3459;%%

%\cite{Fischer:2003rp}
\bibitem{Fischer:2003rp}
  C.~S.~Fischer and R.~Alkofer,
  %``Nonperturbative propagators, running coupling and dynamical quark mass of Landau gauge QCD,''
  Phys.\ Rev.\ D {\bf 67} (2003) 094020
  [hep-ph/0301094].
  %%CITATION = HEP-PH/0301094;%%

%\cite{Bender:1996bb}
\bibitem{Bender:1996bb}
  A.~Bender, C.~D.~Roberts and L.~Von Smekal,
  %``Goldstone theorem and diquark confinement beyond rainbow ladder approximation,''
  Phys.\ Lett.\ B {\bf 380} (1996) 7
  [nucl-th/9602012].
  %%CITATION = NUCL-TH/9602012;%%

%\cite{Watson:2004kd}
\bibitem{Watson:2004kd}
  P.~Watson, W.~Cassing and P.~C.~Tandy,
  %``Bethe-Salpeter meson masses beyond ladder approximation,''
  Few Body Syst.\  {\bf 35} (2004) 129
  [hep-ph/0406340].
  %%CITATION = HEP-PH/0406340;%%

%\cite{Bhagwat:2004hn}
\bibitem{Bhagwat:2004hn}
  M.~S.~Bhagwat, A.~Holl, A.~Krassnigg, C.~D.~Roberts and P.~C.~Tandy,
  %``Aspects and consequences of a dressed quark gluon vertex,''
  Phys.\ Rev.\ C {\bf 70} (2004) 035205
  [nucl-th/0403012].
  %%CITATION = NUCL-TH/0403012;%%

%\cite{Matevosyan:2006bk}
\bibitem{Matevosyan:2006bk}
  H.~H.~Matevosyan, A.~W.~Thomas and P.~C.~Tandy,
  %``Quark-gluon vertex dressing and meson masses beyond ladder-rainbow truncation,''
  Phys.\ Rev.\ C {\bf 75} (2007) 045201
  [nucl-th/0605057].
  %%CITATION = NUCL-TH/0605057;%%

%\cite{Alkofer:2008tt}
\bibitem{Alkofer:2008tt}
  R.~Alkofer, C.~S.~Fischer, F.~J.~Llanes-Estrada and K.~Schwenzer,
  %``The Quark-gluon vertex in Landau gauge QCD: Its role in dynamical chiral symmetry breaking and quark confinement,''
  Annals Phys.\  {\bf 324} (2009) 106
  [arXiv:0804.3042 [hep-ph]].
  %%CITATION = ARXIV:0804.3042;%%

 %\cite{Fischer:2007ze}
\bibitem{Fischer:2007ze}
  C.~S.~Fischer, D.~Nickel and J.~Wambach,
  %``Hadronic unquenching effects in the quark propagator,''
  Phys.\ Rev.\ D {\bf 76} (2007) 094009
  [arXiv:0705.4407 [hep-ph]].
  %%CITATION = ARXIV:0705.4407;%%


 %\cite{Fischer:2009jm}
\bibitem{Fischer:2009jm}
  C.~S.~Fischer and R.~Williams,
  %``Probing the gluon self-interaction in light mesons,''
  Phys.\ Rev.\ Lett.\  {\bf 103} (2009) 122001
  [arXiv:0905.2291 [hep-ph]].
  %%CITATION = ARXIV:0905.2291;%%

%\cite{Fischer:2008wy}
\bibitem{Fischer:2008wy}
C.~S.~Fischer and R.~Williams,
%``Beyond the rainbow: Effects from pion back-coupling,''
Phys.\ Rev.\ D {\bf 78} (2008) 074006
[arXiv:0808.3372 [hep-ph]].

%\cite{Fischer:2005en}
\bibitem{Fischer:2005en}
  C.~S.~Fischer, P.~Watson and W.~Cassing,
  %``Probing unquenching effects in the gluon polarisation in light mesons,''
  Phys.\ Rev.\ D {\bf 72} (2005) 094025
  [hep-ph/0509213].
  %%CITATION = HEP-PH/0509213;%%

%\cite{Chang:2009zb}
\bibitem{Chang:2009zb}
  L.~Chang and C.~D.~Roberts,
  %``Sketching the Bethe-Salpeter kernel,''
  Phys.\ Rev.\ Lett.\  {\bf 103} (2009) 081601
  [arXiv:0903.5461 [nucl-th]].
  %%CITATION = ARXIV:0903.5461;%%

%\cite{Chang:2010hb}
\bibitem{Chang:2010hb}
  L.~Chang, Y.~-X.~Liu and C.~D.~Roberts,
  %``Dressed-quark anomalous magnetic moments,''
  Phys.\ Rev.\ Lett.\  {\bf 106} (2011) 072001
  [arXiv:1009.3458 [nucl-th]].
  %%CITATION = ARXIV:1009.3458;%%

%\cite{Chang:2011ei}
\bibitem{Chang:2011ei}
  L.~Chang and C.~D.~Roberts,
  %``Tracing masses of ground-state light-quark mesons,''
  Phys.\ Rev.\ C {\bf 85} (2012) 052201
  [arXiv:1104.4821 [nucl-th]].
  %%CITATION = ARXIV:1104.4821;%%

%\cite{CJT}
\bibitem{CJT}
  J.~M.~Cornwall, R.~Jackiw and E.~Tomboulis,
  %``Effective Action for Composite Operators,''
  Phys.\ Rev.\ D {\bf 10} (1974) 2428.
  %%CITATION = PHRVA,D10,2428;%%

%\cite{Berges:2004pu}
\bibitem{Berges:2004pu}
  J.~Berges,
  %``N-particle irreducible effective action techniques for gauge theories,''
  Phys.\ Rev.\ D {\bf 70} (2004) 105010
  [hep-ph/0401172].
  %%CITATION = HEP-PH/0401172;%%

\bibitem{Ball:1980ay} 
  J.~S.~Ball and T.~-W.~Chiu,
  %``Analytic Properties of the Vertex Function in Gauge Theories. 1.,''
  Phys.\ Rev.\ D {\bf 22}, 2542 (1980).
  %%CITATION = PHRVA,D22,2542;%%

%\cite{Alkofer:2002bp}
\bibitem{Alkofer:2002bp}
  R.~Alkofer, P.~Watson and H.~Weigel,
  %``Mesons in a Poincare covariant Bethe-Salpeter approach,''
  Phys.\ Rev.\ D {\bf 65} (2002) 094026
  [hep-ph/0202053].
  %%CITATION = HEP-PH/0202053;%%

\bibitem{GellMann:1968rz} 
  M.~Gell-Mann, R.~J.~Oakes and B.~Renner,
  %``Behavior of current divergences under SU(3) x SU(3),''
  Phys.\ Rev.\  {\bf 175}, 2195 (1968).
  %%CITATION = PHRVA,175,2195;%%

\bibitem{Eichten:1974et}
  E.~Eichten and F.~Feinberg,
  %``Dynamical Symmetry Breaking Of Nonabelian Gauge Symmetries,''
  Phys.\ Rev.\  D {\bf 10}, 3254 (1974).
  %%CITATION = PHRVA,D10,3254;%%



%\cite{Maris:1999nt}
\bibitem{Maris:1999nt} 
  P.~Maris and P.~C.~Tandy,
  %``Bethe-Salpeter study of vector meson masses and decay constants,''
  Phys.\ Rev.\ C {\bf 60}, 055214 (1999)
  [nucl-th/9905056].
  %%CITATION = NUCL-TH/9905056;%%



%\cite{Krassnigg:2009gd}
\bibitem{Krassnigg:2009gd} 
  A.~Krassnigg,
  %``Excited mesons in a Bethe-Salpeter approach,''
  PoS CONFINEMENT {\bf 8}, 075 (2008)
  arXiv:0812.3073 [nucl-th].
  %%CITATION = ARXIV:0812.3073;%%

 %\cite{Krassnigg:2009zh}
\bibitem{Krassnigg:2009zh}
  A.~Krassnigg,
  %``Survey of J=0,1 mesons in a Bethe-Salpeter approach,''
  Phys.\ Rev.\ D {\bf 80} (2009) 114010
  [arXiv:0909.4016 [hep-ph]].
  %%CITATION = ARXIV:0909.4016;%%
 
 %\cite{Heupel:2012ua}
\bibitem{Heupel:2012ua}
  W.~Heupel, G.~Eichmann and C.~S.~Fischer,
  %``Tetraquark Bound States in a Bethe-Salpeter Approach,''
  Phys.\ Lett.\ B {\bf 718} (2012) 545
  [arXiv:1206.5129 [hep-ph]].
  %%CITATION = ARXIV:1206.5129;%%
 
\end{thebibliography}
\end{document}